\documentclass[12pt,english]{article}
\pdfoutput=1
\usepackage[T1]{fontenc}
\usepackage{graphicx,array}
\usepackage{cite}
\usepackage{color}
\usepackage{bm}
\usepackage{dsfont}
\usepackage{latexsym}
\usepackage{amsthm}
\usepackage{amsmath}
\usepackage{stackengine}
\usepackage[titletoc]{appendix}
\usepackage{enumitem}
\usepackage{amssymb}
\usepackage{float}
\usepackage[unicode=true,
 bookmarks=true,bookmarksnumbered=false,bookmarksopen=false,
 breaklinks=false,pdfborder={0 0 1},backref=false,colorlinks=true]
 {hyperref}
\hypersetup{pdftitle={Horizon Acoustics of the GHS Black Hole and the Spectrum of AdS2},
 pdfauthor={Achilleas P. Porfyriadis, Grant N. Remmen},
 citecolor=blue,linkcolor=black,urlcolor=black}
\usepackage{breakurl}
\usepackage[hang,flushmargin]{footmisc}

\newcolumntype{L}[1]{>{\raggedright\let\newline\\\arraybackslash\hspace{0pt}}m{#1}}
\newcolumntype{C}[1]{>{\centering\let\newline\\\arraybackslash\hspace{0pt}}m{#1}}

\def\simgt{\mathrel{\lower2.5pt\vbox{\lineskip=0pt\baselineskip=0pt
           \hbox{$>$}\hbox{$\sim$}}}}
\def\simlt{\mathrel{\lower2.5pt\vbox{\lineskip=0pt\baselineskip=0pt
           \hbox{$<$}\hbox{$\sim$}}}}

\newcommand{\be}{\begin{equation}}
\newcommand{\ee}{\end{equation}}
\newcommand{\bea}{\begin{eqnarray}}
\newcommand{\eea}{\end{eqnarray}}
\newcommand{\Eq}[1]{Eq.~(\ref{#1})}
\newcommand{\Sec}[1]{Sec.~\ref{#1}}

\newcommand{\Fig}[1]{Fig.~\ref{#1}}

\newcommand*\oline[1]{%
  \vbox{%
    \hrule height 0.5pt
    \kern0.68ex
    \hbox{%
      \kern-0.1em
      \ifmmode#1\else\ensuremath{#1}\fi
      \kern-0.1em
    }
  }
}

\definecolor{nicered}{rgb}{0.7,0.1,0.1}
\definecolor{nicegreen}{rgb}{0.1,0.5,0.1}
\hypersetup{
 colorlinks,citecolor=black,linkcolor=black,urlcolor=black}
\usepackage{xcolor}

\renewcommand{\Box}{\square}

\begin{document}

\interfootnotelinepenalty=10000
\baselineskip=18pt
\hfill

\vspace{2cm}
\thispagestyle{empty}
\begin{center}
{\LARGE \bf
Horizon Acoustics of the GHS Black Hole\\ \vspace{2mm} and the Spectrum of AdS$_{\boldsymbol{2}}$
}\\
\bigskip\vspace{1cm}{
{\large Achilleas P. Porfyriadis${}^{a,b}$ and Grant N. Remmen${}^{c,d}$}
} \\[7mm]
{
\it ${}^a$Center for the Fundamental Laws of Nature, Harvard University, Cambridge, MA 02138
${}^b$Black Hole Initiative, Harvard University, Cambridge, MA 02138\\[1.5 mm]
${}^c$Kavli Institute for Theoretical Physics, University of California, Santa Barbara, CA 93106
${}^d$Department of Physics, University of California, Santa Barbara, CA 93106 \\[1.5 mm]}
\let\thefootnote\relax\footnote{e-mail: 
\url{porfyr@g.harvard.edu}, \url{remmen@kitp.ucsb.edu}}
 \end{center}

\bigskip
\centerline{\large\bf Abstract}
\begin{quote} \small
We uncover a novel structure in Einstein-Maxwell-dilaton gravity: an ${\rm AdS}_2 \times S^2$ solution in string frame, which can be obtained by a near-horizon limit of the extreme GHS black hole with dilaton coupling $\lambda \neq 1$.
Unlike the Bertotti-Robinson spacetime, our solution has independent length scales for the ${\rm AdS}_2$ and $S^2$,
with ratio controlled by $\lambda$.
We solve the perturbation problem for this solution, finding the independently propagating towers of states in terms of superpositions of gravitons, photons, and dilatons and their associated effective potentials.
These potentials describe modes obeying conformal quantum mechanics, with couplings that we compute, and can be recast as giving the spectrum of the effective masses of the modes.
By dictating the conformal weights of boundary operators, this spectrum provides crucial data for any future construction of a holographic dual to these ${\rm AdS}_2\times S^2$ configurations.

\end{quote}
	
\setcounter{footnote}{0}

\newpage
\tableofcontents
\newpage

\section{Introduction}\label{sec:intro}
Black holes at extremality develop a near-horizon geometry that may be obtained by an appropriate scaling limit. This limit endows the geometry near the horizon with scaling symmetry. For a wide range of theories, a full ${\rm SL}(2)$ isometry group then emerges dynamically, as a consequence of the Einstein equations~\cite{Kunduri:2007vf}. As a result, two-dimensional anti-de Sitter space (${\rm AdS}_2$) is ubiquitous in the near-horizon scaling limits of extreme black holes. This includes the astrophysically relevant case of extreme Kerr~\cite{Bardeen:1999px}.

On the other hand, the extreme GHS black hole~\cite{GHS,Gibbons1982,GibbonsMaeda1988} in Einstein-Maxwell-dilaton theory,
\begin{equation}
{\cal L} =  R - 2\nabla_\mu\phi\nabla^\mu\phi - \frac{1}{2}e^{-2\lambda\phi}F_{\mu\nu}F^{\mu\nu},\label{eq:action}
\end{equation}
is known to defy such odds. Near extremality, the horizon area of the GHS black hole shrinks to zero, causing the extreme black hole to be singular precisely on the horizon. Nevertheless, as soon as the singularity is cured, ${\rm AdS}_2$ makes an appearance in the near-horizon geometry of extreme GHS as well. For example, it was recently shown in Ref.~\cite{Herdeiro:2021gbw} that string-theoretic $\alpha'$ corrections to \Eq{eq:action} give rise to field equations with an exact ${\rm AdS}_2 \times S^2$ solution, and numerical evidence suggests that it indeed represents the near-horizon region of a regular extreme GHS black hole in that theory. 
Similarly, the addition of a particular potential for the dilaton that regularizes the extremal horizon yields an ${\rm AdS}_2 \times S^2$ solution as well \cite{Astefanesei:2019qsg}.
In this paper, we point out that ${\rm AdS}_2 \times S^2$ arises more directly in the theory \eqref{eq:action} by switching to the string frame. In string frame, the GHS black hole is regular at extremality too, and we show that ${\rm AdS}_2 \times S^2$ is indeed obtained via a scaling limit of extreme GHS.

Interestingly, our ${\rm AdS}_2 \times S^2$ solution in the string-frame Einstein-Maxwell-dilaton theory has a dilaton that breaks the full ${\rm SL}(2)$ symmetry, so that our near-horizon isometry group is only enhanced by scaling symmetry compared to GHS. We find that the scaling symmetry is sufficient to ensure that perturbations are governed by the Schr\"odinger equation of conformal quantum mechanics. 
Nevertheless, the ${\rm SL}(2)$ invariance of the gravitoelectromagnetic sector appears to be responsible for the surprising fact that we find graviton and photon modes separate cleanly from the dilaton modes, so that two of our master variables are constructed from the same perturbation components as in the $\lambda=0$ Einstein-Maxwell theory.

${\rm AdS}_2$ holography has been recognized from the early days of AdS/CFT to pose unique challenges compared to its higher dimensional analogues~\cite{Strominger:1998yg, Maldacena:1998uz}. Recently, beginning with Refs.~\cite{Almheiri:2014cka, Jensen:2016pah, Maldacena:2016upp, Engelsoy:2016xyb}, there has been progress towards understanding the universal aspects of ${\rm AdS}_2$ holography that are relevant for the s-wave sector of gravitational dynamics near extreme black hole horizons, as described by two-dimensional Jackiw-Teitelboim gravity.
Going beyond the s-wave sector will be less universal. In particular, the holographic dictionary will include operators with conformal dimensions that are fixed by the Kaluza-Klein (KK) spectrum of ${\rm AdS}_2$ specific to the theory and black hole whose near-horizon it describes. In pure Einstein-Maxwell theory, the ${\rm AdS}_2$ spectrum in the Bertotti-Robinson solution may be found in Refs.~\cite{Porfyriadis:2018yag, Porfyriadis:2018jlw} and \Sec{sec:RN}. However, in Bertotti-Robinson the ${\rm AdS}_2$ radius is equal to that of the $S^2$, and this leads to difficulties in the effective field theory interpretation of KK modes. On the other hand, the ${\rm AdS}_2 \times S^2$ solution in Einstein-Maxwell-dilaton theory  identified in this paper has independent radii for the ${\rm AdS}_2$ and $S^2$ factors, with a ratio controlled by $\lambda$. Therefore, our ${\rm AdS}_2$ bulk solution here is a smoother arena for studying ${\rm AdS}_2$ holography beyond the s-wave sector. The complete solution to the perturbation problem we present in this paper writes crucial entries of the holographic dictionary; namely, we find the conformal weights of operators dual to the bulk metric, gauge, and dilaton fields.

This paper is organized as follows. In \Sec{sec:AdS}, we present the new ${\rm AdS}_2 \times S^2$ solution to the string-frame equations of motion of Einstein-Maxwell-dilaton gravity and show how this solution can be obtained via a near-horizon scaling limit of the extreme GHS black hole.

\begin{figure}[t]
\begin{center}
\includegraphics[width=12cm]{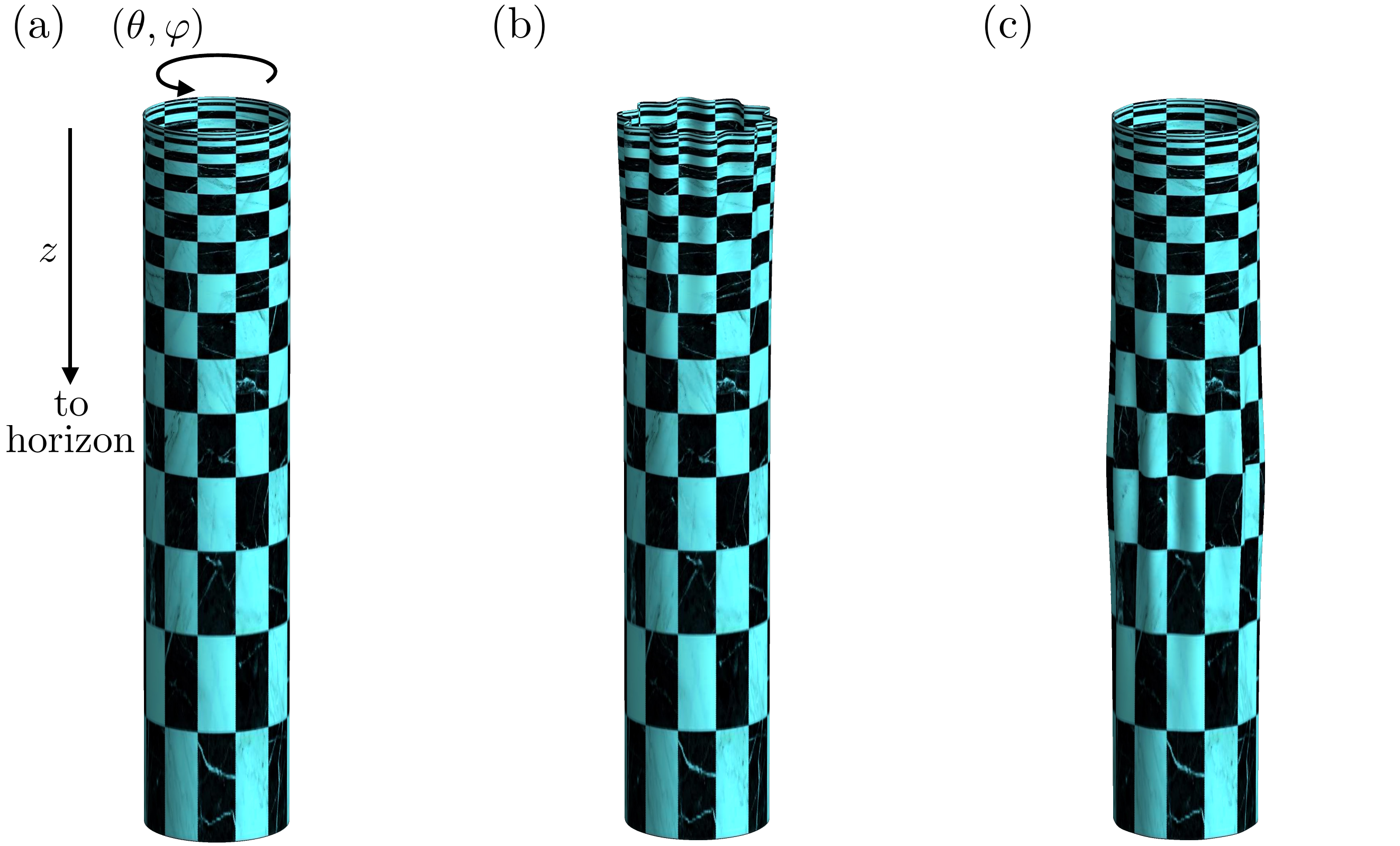}
\end{center}
\vspace{-0.5cm}
\caption{Illustration of spatial slices of our ${\rm AdS}_2 \times S^2$ solution and perturbations, with one of the two angular directions suppressed. (a) The unperturbed solution, with boundary at $z=0$ and Poincar\'e horizon at $z=\infty$. (b) A perturbation near the boundary. (c) A perturbation deeper in the bulk.
We will fully categorize the propagating modes, solving the Regge-Wheeler-Zerilli problem in this geometry and diagonalizing the mass matrix.
}
\label{fig:AdS2}
\end{figure}

We then turn to the study of perturbations around this ${\rm AdS}_2\times S^2$ solution.
As a major result of this work, we identify all the linear modes for propagating solutions in the throat as depicted in \Fig{fig:AdS2}---constructed out of the metric, gauge field, and dilaton degrees of freedom---and calculate the associated effective potentials.
We find that the equation of motion for these modes is always the time-independent Schr\"odinger equation of conformal quantum mechanics, as a consequence of the scaling symmetry of the ${\rm AdS}_2$, with potential $\propto g/z^2$ for Poincar\'e coordinate $z$.
The parameter $g$ encodes the effective mass, or equivalently conformal weight, of the mode solutions.
In \Sec{sec:summary} we summarize the results of our calculation, explicitly giving the spectra of modes for our ${\rm AdS}_2 \times S^2$ solution.

In \Sec{sec:linearized} we derive the linearized equations of motion for perturbative Einstein-Maxwell-dilaton theory in string frame, as well as detail the Regge-Wheeler-Zerilli ansatz we will use, in both the axial and polar cases, for the form of the graviton, photon, and scalar perturbations.
We present the details of the calculation itself for the axial and polar modes in Secs.~\ref{sec:axial} and \ref{sec:polar}, respectively, obtaining the appropriate admixtures of the components of the metric, gauge field, and dilaton that propagate independently and finding the effective masses for multipoles with $\ell \geq 2$.
We treat the special cases of $\ell = 1$ and $\ell = 0$ modes in Secs.~\ref{sec:l1} and \ref{sec:l0}, respectively.

We discuss special topics in \Sec{sec:discussion}, specifically, 1) the functional form of the propagating solutions to our equations of motion; 2) the $\lambda = \sqrt{3}$ case, which we show arises from five-dimensional general relativity as the KK reduction of an extra dimension as a Hopf fiber; and 3) the $\lambda = 0$ case of the near-horizon Reissner-Nordstr\"om black hole, generalized to accommodate dyonic charges, where we find a ${\rm U}(1)\times \mathbb{Z}_2$ symmetry in the effective potential.
We conclude and discuss future directions in \Sec{sec:conclusions}.

\section{AdS and GHS}\label{sec:AdS}

Conformally transforming the Lagrangian~\eqref{eq:action} in terms of the string-frame metric $\widetilde g_{\mu\nu}= e^{2\lambda\phi} g_{\mu\nu}$ (and subsequently dropping explicit tildes), we have
\be
{\cal L}_{\rm string} =  e^{-2\lambda\phi}\left[R+2(3\lambda^{2}-1)\nabla_\mu\phi\nabla^\mu\phi-\frac{1}{2}F_{\mu\nu}F^{\mu\nu}\right],\label{eq:stringaction}
\ee
up to a total derivative.
In string frame, the Einstein equation is
\be
\begin{aligned}
R_{\mu\nu}-\frac{1}{2}Rg_{\mu\nu}- F_{\mu\rho}F_\nu^{\;\;\rho}+\frac{1}{4}g_{\mu\nu}F_{\rho\sigma}F^{\rho\sigma}-2(1 - \lambda^2)\nabla_{\mu}\phi\nabla_{\nu}\phi \\ +(1 + \lambda^2)(\nabla\phi)^{2}g_{\mu\nu}+2\lambda\left(\nabla_{\mu}\nabla_{\nu}\phi - g_{\mu\nu} \Box\phi \right)& =0,
\end{aligned}\label{eq:eomstring1}
\ee
while the Maxwell and Klein-Gordon equations are, respectively,
\be 
\nabla_{\mu}F^{\mu\nu} -2\lambda F^{\mu\nu}\nabla_{\mu}\phi =0\label{eq:eomstring2}
\ee
and
\be 
2(3\lambda^{2}-1)\Box\phi+\lambda\left[R-2(3\lambda^{2}-1)(\nabla\phi)^{2}-\frac{1}{2}F_{\mu\nu}F^{\mu\nu}\right]  =0.
\label{eq:eomstring3}
\ee

For arbitrary $\lambda \neq 1$, we find that the equations of motion in Eqs.~\eqref{eq:eomstring1}, \eqref{eq:eomstring2}, and \eqref{eq:eomstring3} admit a novel magnetically charged ${\rm AdS}_2\times S^2$ solution:
\be 
\boxed{
	\begin{aligned}
		({\rm d}s^{2})_{{\rm string}}&=\left(\frac{1+\lambda^{2}}{1-\lambda^2}\right)^{2}\frac{r_{0}^{2}}{z^2}(-{\rm d}t^{2}+{\rm d}z^{2})+r_{0}^{2}{\rm d}\Omega^{2}\\
		F &= p\sin \theta\, {\rm d}\theta\wedge {\rm d}\varphi \\
		\phi &= \frac{\lambda}{1-\lambda^2}\log z + \phi_0,
	\end{aligned}
}\label{eq:metstring}
\ee
where we have defined the constants
\be
\begin{aligned}
p &= m\sqrt{2(1+\lambda^2)}\\
r_0 &= m(1+\lambda^2)
\end{aligned}\label{eq:defconstants}
\ee
and where $\phi_0$ and $m$ are arbitrary.
This solution is reminiscent of the magnetic Bertotti-Robinson ${\rm AdS}_2 \times S^2$ obtained from the near-horizon limit of an extreme Reissner-Nordstr\"om black hole, and indeed \Eq{eq:metstring} reduces to Bertotti-Robinson for $\lambda=0$.
However, unlike Bertotti-Robinson, where the AdS length scale and radius of the two-sphere are inextricably equated, the solution in \Eq{eq:metstring} has the favorable feature that these two length scales are independent, with a tunable ratio parameterized by the dilaton coupling.
With $r_0$ giving the radius of the $S^2$, the ${\rm AdS}_2$ length scale for our solution is given by
\be 
r_{\rm AdS}^2 = r_0^2 \left(\frac{1+\lambda^2}{1-\lambda^2}\right)^2.\label{eq:rAdS}
\ee
A remarkable aspect of the solution in \Eq{eq:metstring} is that the dilaton is not constant, and it therefore breaks the full ${\rm SL}(2)$ symmetry associated with the ${\rm AdS}_2$ factor. Specifically, \Eq{eq:metstring} exhibits scaling symmetry, but the special conformal transformation does not leave the dilaton invariant.
In addition, we note that the dilaton diverges on the Poincar\'e horizon of ${\rm AdS}_2$.

Our ${\rm AdS}_2 \times S^2$ solution in \Eq{eq:metstring} can be interpreted as a near-horizon limit of the GHS black hole~\cite{GHS} in string frame.
In Einstein frame, the extreme GHS black hole is given by
\be
\begin{aligned}
({\rm d}s^2)_{\rm GHS, Einstein} &=  -\left(1-\frac{r_{0}}{r}\right)^{\frac{2}{1+\lambda^{2}}}{\rm d}t^{2}+\left(1-\frac{r_{0}}{r}\right)^{-\frac{2}{1+\lambda^{2}}}{\rm d}r^{2}+r^{2}\left(1-\frac{r_{0}}{r}\right)^{\frac{2\lambda^{2}}{1+\lambda^{2}}}{\rm d}\Omega^{2}\\
\phi &= -\frac{\lambda}{1+\lambda^{2}}\log\left(1-\frac{r_{0}}{r}\right),
\end{aligned}
\ee
with $F$, $p$, and $r_0$ as in Eqs.~\eqref{eq:metstring} and \eqref{eq:defconstants}.
We can generalize by adding an arbitrary constant $\phi_0$ to the dilaton, which then requires either rescaling the magnetic charge $p$ by $e^{\lambda\phi_0}$ or leaving $p$ fixed but rescaling the Einstein-frame metric by $e^{-2\lambda\phi_0}$.
We will opt for the latter choice, as it allows us to write the string-frame metric independently of $\phi_0$.

For $\lambda \neq 0$, the extreme GHS solution possesses a singular, zero-area horizon at $r=r_0$.
In string frame,
\be
({\rm d}s^2)_{\rm GHS,string} = -\left(1-\frac{r_{0}}{r}\right)^{\frac{2(1-\lambda^{2})}{1+\lambda^{2}}}{\rm d}t^{2}+\left(1-\frac{r_{0}}{r}\right)^{-2}{\rm d}r^{2}+r^{2}{\rm d}\Omega^{2} ,\label{eq:GHSstringmetric}
\ee
the area is finite. 
For dilaton coupling $|\lambda| \geq 1$ one finds that the string-frame metric is in fact horizonless.
For $|\lambda|<1$, the horizon at $r=r_0$ is nonsingular in the string frame metric~\eqref{eq:GHSstringmetric}.
This can be seen by computing the Riemann tensor and transforming to a local Lorentz frame, where one finds that all components are finite in the $r\rightarrow r_0$ limit.

Another way to see that the $r=r_0$ surface represents a nonsingular horizon for $|\lambda|<1$ is to define an analogue of Eddington-Finkelstein coordinates. We first define a radial coordinate $\tilde r = r_0\,B\left(1-\frac{r_0}{r};\frac{1-\lambda^2}{1+\lambda^2},-1\right)$, where $B(x;a,b)=\int_0^x y^{a-1}(1-y)^{b-1}{\rm d} y$ is the incomplete Euler beta function, in terms of which we have ${\rm d}\tilde r^2/f(r) = {\rm d}r^2/\left(1-\frac{r_0}{r}\right)^2$ for $f(r)=\left(1-\frac{r_0}{r}\right)^{\frac{2(1-\lambda^2)}{1+\lambda^2}}$.
Writing the inverse as $r(\tilde r)$ and denoting $f(r(\tilde r))$ as $\tilde f(\tilde r)$, we define a tortoise coordinate in the usual manner, ${\rm d}r_* = {\rm d}\tilde r/\tilde f(\tilde r)$.
In terms of $v=t+r_*$, the metric in \Eq{eq:GHSstringmetric} now becomes $\left({\rm d}s^2\right)_{\rm GHS,string} = -\tilde f(\tilde r)\,{\rm d}v^2 +2\, {\rm d}v\,{\rm d}\tilde r + [r(\tilde r)]^2{\rm d}\Omega^2$ which is manifestly nonsingular at $r=r_0$.
It should be noted however that the extension of the string-frame metric past the horizon using the above coordinates is not infinitely differentiable. Rather, for general $\lambda \in (0,1)$ the metric is $C^1$ at the horizon.\footnote{For general $\lambda \in (0,1)$ we find that $r(\tilde r)$ (and hence $g_{\theta\theta}$) is $C^1$ and $g_{vv} = -\tilde f(\tilde r)$ is $C^3$. That said, as $\lambda$ approaches unity, the degree of differentiability grows. For example, requiring $1/3 < \lambda^2 < 1$ gives us $C^2$ for $r(\tilde{r})$ and $C^4$ for $\tilde f(\tilde r)$.  We thank the referee for emphasizing this point to us.}
It is also worth noting that what happened here is that
the $r=r_0$ singularity in the Einstein-frame geometry has been absorbed entirely by the Weyl rescaling with the dilaton profile, which itself is singular at the horizon.

Let us define the near-horizon limit as follows. 
We set
\be 
\begin{aligned}
t & \rightarrow r_0 \epsilon^{-\frac{1-\lambda^2}{1+\lambda^{2}}}\hat t\\
r & \rightarrow r_{0}(1+\epsilon \hat r)\\
\phi_{0} & \rightarrow\hat{\phi}_{0}+\frac{\lambda}{1+\lambda^{2}}\log\epsilon.
\end{aligned}\label{eq:scaling}
\ee
Then the GHS solution becomes:
\be 
\begin{aligned}
\left({\rm d}s^{2}\right)_{{\rm string}} & =r_0^2 \left[-\left(\frac{1}{\hat r}+\epsilon\right)^{-\frac{2(1-\lambda^2)}{1+\lambda^{2}}}{\rm d}\hat t^{2}+\left(\frac{1}{\hat r}+\epsilon\right)^{2}{\rm d}\hat r^{2}+(1+\epsilon \hat r)^{2}{\rm d}\Omega^{2}\right]\\
\phi & =\hat{\phi}_{0}+\frac{\lambda}{1+\lambda^{2}}\log\left(\frac{1}{\hat r}+\epsilon\right).
\end{aligned}\label{eq:solepsilon}
\ee
The $\epsilon\rightarrow 0$ limit of Eq.~\eqref{eq:solepsilon} is regular:
\be 
\begin{aligned}
\left({\rm d}s^{2}\right)_{{\rm string}} & =r_0^2 \left(-\hat r^{\frac{2(1-\lambda^2)}{1+\lambda^{2}}}{\rm d}\hat t^{2}+\frac{{\rm d}\hat r^{2}}{\hat r^2} +{\rm d}\Omega^{2}\right)\\
\phi & =\hat{\phi}_{0}-\frac{\lambda}{1+\lambda^{2}}\log \hat r.
\end{aligned}
\ee
Let us then perform a coordinate transformation,
\be 
\hat r^{\frac{1-\lambda^2}{1+\lambda^2}} = \left|\frac{1+\lambda^{2}}{1-\lambda^2}\right|\frac{1}{z}
,\label{eq:coord}
\ee
under which, after setting $\hat{\phi}_{0}=\phi_0 + \frac{\lambda}{1-\lambda^2}\log\left|\frac{1+\lambda^{2}}{1-\lambda^2}\right|$ and relabeling $\hat t$ as $t$, we find our ${\rm AdS}_{2}\times S^{2}$ solution precisely as given in \Eq{eq:metstring}.

For $\lambda < 1$, the horizon of the original GHS black hole has become the Poincar\'e horizon at $z\rightarrow \infty$, while the AdS boundary at $z = 0$ corresponds roughly to the region where the near-horizon throat joins the asymptotically flat geometry.
For $\lambda > 1$, the roles of $z=\infty$ and $z=0$ are reversed, in which case the region very close to but outside the Einstein-frame horizon of the GHS black hole is mapped to the region near the ${\rm AdS}_2$ boundary of a Poincar\'e patch.
For $\lambda = 1$---which corresponds to the low-energy effective action of the heterotic string---the solution does not exist, since for dilaton coupling equal to unity the scaling~\eqref{eq:scaling} and coordinate transformation~\eqref{eq:coord} are singular.

\section{Summary of spectra}\label{sec:summary}

We now wish to study perturbation theory around the ${\rm AdS}_2 \times S^2$ solution presented in \Sec{sec:AdS}.
Expanding the equations of motion in Eqs.~\eqref{eq:eomstring1}, \eqref{eq:eomstring2}, and \eqref{eq:eomstring3} to linear order in perturbations of the string-frame metric $\delta g_{\mu\nu}$, gauge field strength $\delta F_{\mu\nu}$, and dilaton $\delta\phi$, we will diagonalize the system and extract the wave functions and associated master equations.
The calculation is the Einstein-Maxwell-dilaton analogue, for our ${\rm AdS}_2 \times S^2$ solution, of the classic calculations of Regge and Wheeler~\cite{ReggeWheeler} and Zerilli~\cite{Zerilli, ZerilliRN}.
The propagating modes are given by superpositions of the graviton, gauge field, and dilaton; this is to be expected, since all three types of fields are nonzero for this background, leading to mixed wave functions.

We will give the explicit expressions for the wave functions, as well as their derivation from the equations of motion, in Secs.~\ref{sec:axial} and \ref{sec:polar}, but for now let us summarize some key results.
Factoring off the angular dependence and time dependence $\propto e^{i\omega t}$, the master variables are functions of the Poincar\'e $z$-coordinate. 
In all cases, we find that the master equation for a given mode $\psi(z)$ satisfies the time-independent Schr\"odinger equation for conformal quantum mechanics,
\be
\psi'' + \omega^2 \psi - \frac{g}{z^2}\psi = 0.\label{eq:conformalQM}
\ee
This form of the potential is a consequence of the scaling symmetry of the ${\rm AdS}_2 \times S^2$ solution.
In terms of the $\Box$ operator defined with respect to the background string metric, the master equation~\eqref{eq:conformalQM} can be recast as a simple wave equation for $\Psi(t,z)  = e^{i\omega t}\psi(z)$,
\be
(\Box  - M^2)\Psi = 0,
\ee
where the mass is fixed by the potential coupling $g$,
\be
M^2 = \frac{g}{r_{\rm AdS}^2}, \label{eq:massg}
\ee
for $r_{\rm AdS}$ given in \Eq{eq:rAdS}. 

We organize the modes by parity and expand in spherical harmonics $Y_{\ell m}(\theta,\varphi)$.
Throughout, we will make use of the spherical symmetry of the background solution to set the azimuthal quantum number to zero without loss of generality, so that the spherical harmonics are replaced with Legendre polynomials of $\cos\theta$.
We use the term ``polar'' (respectively, ``axial'') for modes that have graviton perturbations of even (respectively, odd) parity, i.e., that pick up a sign $(-1)^\ell$ (respectively, $(-1)^{\ell+1}$) under a parity inversion.
For the gauge field, because our background is magnetic and hence odd under parity, the ``polar'' and ``axial'' labels apply in the opposite sense: polar for parity-odd perturbations ($(-1)^{\ell+1}$) and axial for parity-even perturbations ($(-1)^{\ell}$).
The dilaton mode, being parity-even, contributes only in the polar case.
With this identification, the polar and axial modes decouple from each other.\footnote{In the literature, the labels ``electric'' and ``magnetic'' are often applied to the parity-even and -odd modes of the graviton, respectively, but we eschew such identification here to avoid confusion with the gauge field itself (whose perturbation gives rise to both electric and magnetic fields in either case).}
See \Sec{sec:linearized} for explicit expressions for the perturbative  ansatz.

\begin{figure}[H]
\begin{center}
\includegraphics[width=12cm]{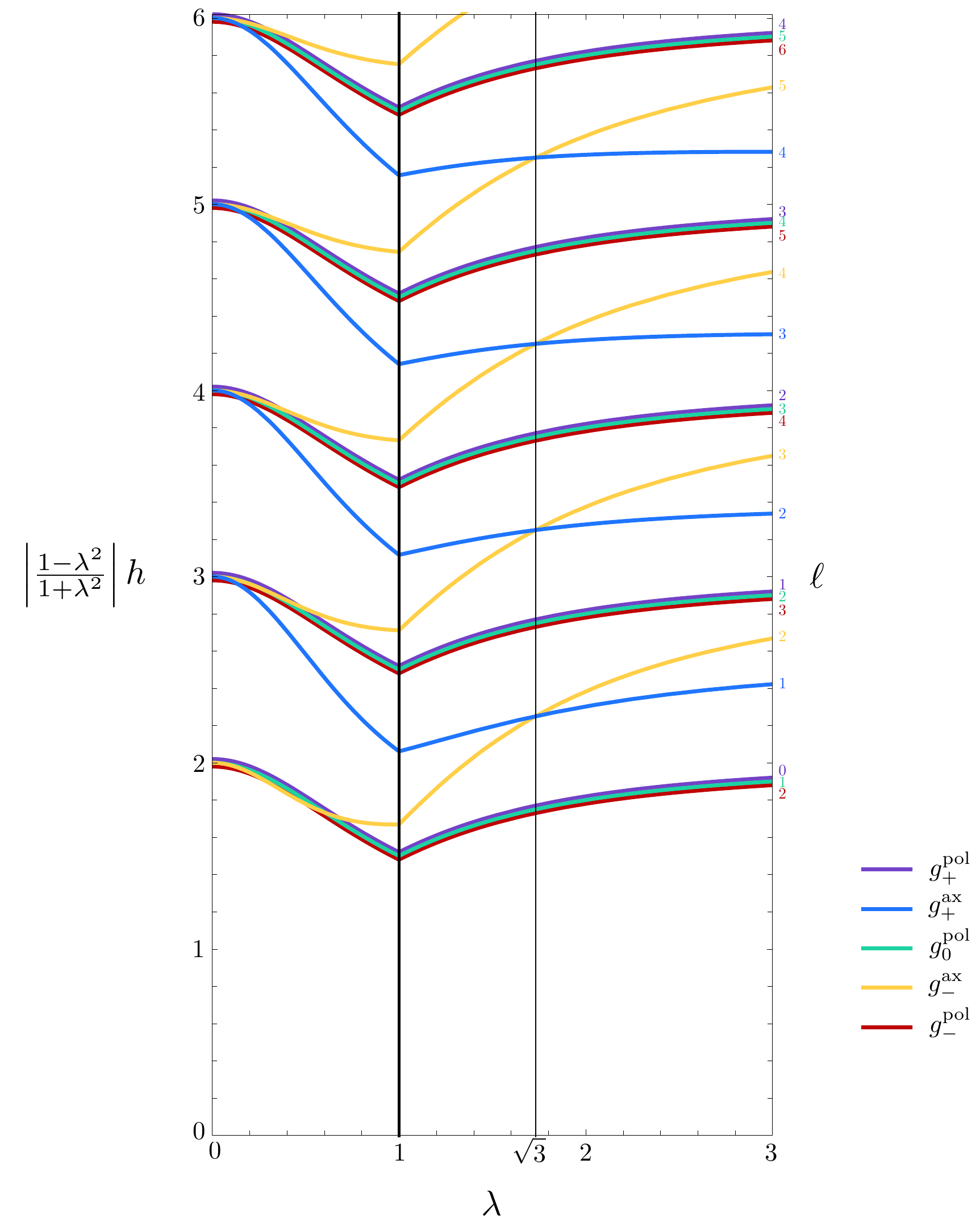}
\end{center}
\vspace{-5mm}
\caption{Spectrum of perturbations around the ${\rm AdS}_2\times S^2$ string-frame solution~\eqref{eq:metstring} in Einstein-Maxwell-dilaton theory. The conformal effective potential $g/z^2$ corresponds to an effective mass term~\eqref{eq:massg}.
For general dilaton coupling $\lambda$, we find five towers of massive states, indexed by angular momentum $\ell \geq 2$, while we find one state for $\ell=0$ and three for $\ell=1$.
The couplings $g$, given in \Eq{eq:gs}, are recast as a conformal weight $h$ via \Eq{eq:weight} and plotted above for various values of $\lambda$, with the corresponding $\ell$ value given at right.
}
\label{fig:spectrum}
\end{figure}

For all $\ell \geq 2$, we find that there are five towers of massive states, of which two are axial and three are polar, with potential couplings $g^{\rm ax}_+$, $g^{\rm ax}_-$, $g^{\rm pol}_+$, $g^{\rm pol}_0$, and $g^{\rm pol}_-$, each indexed by $\ell$:
\be 
\boxed{
\begin{aligned}
g_\pm^{\rm ax} &= \frac{1-\lambda^{2}+\lambda^{4}+\ell(\ell+1)(1+\lambda^{2})^{2}\pm(1+\lambda^{2})\sqrt{4\ell(\ell+1)(1+\lambda^2) + (1-\lambda^2)^2}}{(1-\lambda^2)^{2}} \\
g^{\rm pol}_{-} & =\frac{[(1+\lambda^2)(\ell-1)+1][\lambda^2\ell+\ell-1]}{(1-\lambda^2)^{2}}\\
g^{\rm pol}_{0}&=\frac{[(1+\lambda^2) \ell+1][\lambda^2 (\ell+1)+\ell]}{(1-\lambda^2)^{2}}\\
g^{\rm pol}_{+} & =\frac{[(1+\lambda^{2})(\ell+1)+1][\lambda^{2}(\ell+2)+\ell+1]}{(1-\lambda^2)^{2}}.
\end{aligned}
}\label{eq:gs}
\ee
This counting is to be expected, given the five degrees of freedom present among the graviton, gauge field, and dilaton. 
The $\ell=0$ and $\ell=1$ cases must be treated separately, and we find that there is only one propagating $\ell=0$ mode, with coupling $g^{\rm pol}_+(\ell=0)$, and three $\ell=1$ modes, with couplings $g^{\rm pol}_+(\ell=1)$, $g^{\rm pol}_0(\ell=1)$, and $g^{\rm ax}_+(\ell=1)$.
Finding these couplings---that is, identifying the spectrum of perturbations around our ${\rm AdS}_2 \times S^2$---is a primary result of this work.
This spectrum would inform any attempt to construct a holographic boundary theory dual. 
In terms of the couplings $g$, the conformal weights of the dual operators, depicted in \Fig{fig:spectrum}, are fixed by
\be 
h = \frac{1}{2} + \sqrt{\frac{1}{4}+g},\label{eq:weight}
\ee
so that $g = M^2 r_{\rm AdS}^2 = h(h-1)$ ~\cite{Witten:1998qj, Gubser:1998bc}.
For example, for massive free fields propagating on an ${\rm AdS}_2$ bulk, the  highest-weight state for the field corresponds, with $h$ an integer, to a primary operator in the one-dimensional boundary theory obtained via $h$ normal derivatives acting on the field, $(n^\mu \partial_\mu)^h \Psi$~\cite{Strominger:1998yg, Spradlin:1999bn}.
An understanding of the conformal weights \eqref{eq:weight} of the graviton/photon/dilaton perturbations, solved for in this paper, is of crucial importance to the eventual goal of constructing a holographic dual to our customizable ${\rm AdS}_2 \times S^2$ bulk.

\begin{figure}[t]
\begin{center}
\includegraphics[width=16cm]{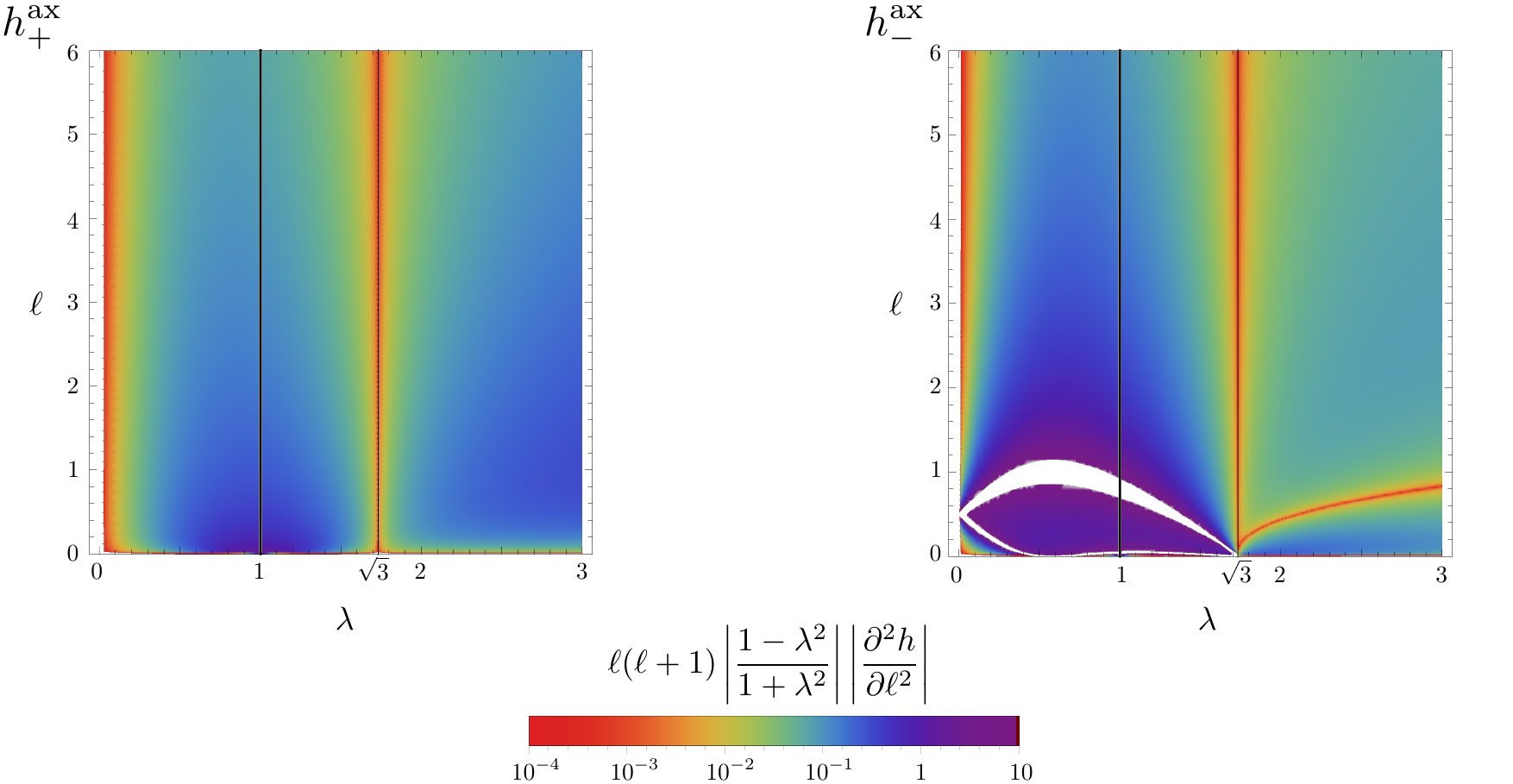}
\end{center}
\vspace{-0.5cm}
\caption{Warping of the tower of conformal weights $h_\pm^{\rm ax}$ for the two axial modes perturbing our ${\rm AdS}_2 \times S^2$ background in Einstein-Maxwell-dilaton theory, corresponding to $g_{\pm}^{\rm ax}$ in \Eq{eq:gs} via \Eq{eq:weight}. We find that a linear tower of axial modes occurs only for $\lambda^2 = 0$ or $3$ (red vertical lines), corresponding respectively either to four-dimensional Einstein-Maxwell theory or to the KK reduction of five-dimensional Einstein gravity.
}
\label{fig:deformation}
\end{figure}

Importantly, the couplings in Eq.~\eqref{eq:gs} are {\it not} simply those of the KK tower for a free particle propagating in the ${\rm AdS}_2 \times S^2$ background, which would simply give $g = \ell(\ell+1)$. Instead, they possess interesting, nontrivial structure.
The conformal weights of the polar modes are linear in $\ell$ for arbitrary $\lambda$. 
However, for the axial modes, the tower is warped for general $\lambda$, and one obtains conformal weights linear in $\ell$ only for the special values of $\lambda = 0$ (corresponding to Einstein-Maxwell) and $\lambda = \sqrt{3}$ (corresponding to KK reduction of Einstein gravity in five dimensions); see \Fig{fig:deformation} for an illustration.

For these special values, the spectra take the elegant forms:\footnote{We write the $\lambda = 0$ case as a limit $\lambda \rightarrow 0$ above, since for $\lambda$ strictly vanishing, the dilaton mode decouples entirely and the number of physical modes will be different; we will consider the case of Einstein-Maxwell theory in more detail in Sec.~\ref{sec:RN}.}
\be
\begin{aligned}
\lambda \rightarrow 0: & \;& (h_-^{\rm pol}, h_-^{\rm ax}, h_0^{\rm pol}, h_+^{\rm ax}, h_+^{\rm pol}) &= (\ell,\ell,\ell+1,\ell+2,\ell+2)\\
\lambda = \sqrt{3}: & &(h_-^{\rm pol}, h_-^{\rm ax}, h_0^{\rm pol}, h_+^{\rm ax}, h_+^{\rm pol}) &= \left(2\ell-\tfrac{1}{2},2\ell+\tfrac{1}{2},2\ell+\tfrac{3}{2},2\ell+\tfrac{5}{2},2\ell+\tfrac{7}{2}\right).
\end{aligned} \label{eq:spectrumsimple}
\ee
Keeping track of which modes apply in the $\ell=0,1$ cases, we find that,  for all $\lambda$ and $\ell$, the modes are always massive, and no massless modes or tachyonic instabilities appear.
See \Fig{fig:spectrum} for details.
The polar modes satisfy $g_-^{\rm pol}(\ell+1) = g_0^{\rm pol}(\ell) = g_+^{\rm pol}(\ell-1)$, leading to a threefold degeneracy in the spectrum for arbitrary $\lambda$.
For $\lambda = \sqrt{3}$, one further has $g_-^{\rm ax}(\ell+1) = g_+^{\rm ax}(\ell)$, leading to a twofold degeneracy in the spectrum among the axial modes.
In the $\lambda\rightarrow 0$ limit, there is a fourfold degeneracy of the lowest-lying state, and generically fivefold degeneracy, as we see in \Eq{eq:spectrumsimple} and \Fig{fig:spectrum}.
Finally, for $\lambda\rightarrow 1$, the masses of all of the towers diverge, as another consequence of the singularity in the ${\rm AdS}_2 \times S^2$ metric when $\lambda = 1$.

\section{Perturbative Einstein-Maxwell-dilaton theory}\label{sec:linearized}

Let us perturb the string-frame equations of motion, sending the string-frame metric $g_{\mu\nu}\rightarrow g_{\mu\nu} + \delta g_{\mu\nu}$, the gauge field strength $F_{\mu\nu} \rightarrow F_{\mu\nu} + \delta F_{\mu\nu}$, and the dilaton $\phi \rightarrow \phi + \delta \phi$.
For brevity of notation, let us write $\delta g_{\mu\nu}$ as $h_{\mu\nu}$, $\delta F_{\mu\nu} = \partial_\mu \delta A_\nu -\partial_\nu \delta A_\mu$ as $f_{\mu\nu} = \partial_\mu a_\nu - \partial_\nu a_\mu$, and $\delta \phi$ as $\chi$, and write the Einstein, Maxwell, and Klein-Gordon equations given in Eqs.~\eqref{eq:eomstring1},~ \eqref{eq:eomstring2},~and~\eqref{eq:eomstring3} as $E_{\mu}^{\;\;\nu} = 0$, $M^\mu = 0$, and $S = 0$, respectively, where for convenience we will raise an index on the Einstein equation.
Expanding these equations of motion to first order in the perturbations and setting the backgroud to our ${\rm AdS}_2 \times S^2$ solution, we have the linearized equations $\delta E_\mu^{\;\;\nu} = 0$, $\delta M^\mu = 0$, and $\delta S = 0$.
We find that these expressions simplify if we instead consider the following equivalent equations of motion, obtained by subtracting off traces and/or components proportional to the background equations: 
\be 
\begin{aligned}
\delta \hat E_{\mu}^{\;\;\nu} &= \delta E_\mu^{\;\;\nu} - \frac{1}{2}\delta^\nu_\mu \delta E_\rho^{\;\;\rho}  + \left(E_\mu^{\;\;\rho} - \frac{1}{2}\delta_\mu^\rho E_\sigma^{\;\;\sigma}\right) h_\rho^{\;\;\nu} \\
\delta \hat M^\mu &= \delta M^\mu + M^\nu h_\nu^{\;\;\mu} \\
\delta \hat S &= \delta S.
\end{aligned}
\ee
We have the linearized Einstein equation $\delta\hat E_\mu^{\;\;\nu} = 0$, where
\be
\begin{aligned}
\delta \hat E_{\mu}^{\;\;\nu} & =\frac{1}{2}\left(\nabla_{\rho}\nabla_{\mu}h^{\rho\nu}+\nabla^{\rho}\nabla^{\nu}h_{\rho\mu}-\Box h_{\mu}^{\;\;\nu}-\nabla_{\mu}\nabla^{\nu}h_{\rho}^{\;\;\rho}\right)\\
 & \qquad+(2\lambda^{2}-2)\left(\nabla_{\mu}\phi\nabla^{\nu}\chi+\nabla_{\mu}\chi\nabla^{\nu}\phi\right)+2\lambda\nabla_{\mu}\nabla^{\nu}\chi\\
 & \qquad-\lambda\nabla_{\rho}\phi\left(\nabla_{\mu}h^{\nu\rho}+\nabla^{\nu}h_{\mu}^{\;\;\rho}-\nabla^{\rho}h_{\mu}^{\;\;\nu}\right)+\lambda h_{\mu}^{\;\;\nu}\left[\Box\phi-2\lambda(\nabla\phi)^{2}\right]\\
 & \qquad+\lambda\delta_{\mu}^{\nu}\left[-h^{\rho\sigma}\nabla_{\rho}\nabla_{\sigma}\phi-\nabla^{\sigma}\phi\left(\nabla^{\rho}h_{\rho\sigma}-\frac{1}{2}\nabla_{\sigma}h_{\rho}^{\;\;\rho}\right)+2\lambda h^{\rho\sigma}\nabla_{\rho}\phi\nabla_{\sigma}\phi\right]\\
  & \qquad+\lambda\delta_{\mu}^{\nu}\left(\Box\chi-4\lambda\nabla^{\rho}\phi\nabla_{\rho}\chi\right)\\
 & \qquad-\frac{1}{2}\delta_{\mu}^{\nu}F_{\rho\alpha}F_{\sigma}^{\;\;\alpha}h^{\rho\sigma}+F_{\mu\rho}F^{\nu\sigma}h_{\;\;\sigma}^{\rho}+\frac{1}{4}F_{\rho\sigma}F^{\rho\sigma}h_{\mu}^{\;\;\nu}-F_{\mu\rho}f^{\nu\rho}-F^{\nu\rho}f_{\mu\rho}+\frac{1}{2}\delta_{\mu}^{\nu}F^{\rho\sigma}f_{\rho\sigma},
\end{aligned}\label{eq:ein}
\ee
the linearized Maxwell equation $\delta \hat M^\mu = 0$, where 
\be
\begin{aligned}
\delta \hat M^{\mu} & =\nabla_{\nu}f^{\nu\mu}-2\lambda f^{\nu\mu}\nabla_{\nu}\phi-h^{\nu\rho}\nabla_{\rho}F_{\nu}^{\;\;\mu}-F_{\rho}^{\;\;\mu}\nabla_{\nu}h^{\nu\rho}-F^{\nu\rho}\nabla_{\nu}h_{\rho}^{\;\;\mu}\\
 & \qquad+\frac{1}{2}F^{\nu\mu}\nabla_{\nu}h_{\rho}^{\;\;\rho}+2\lambda F^{\nu\mu}h_{\nu\rho}\nabla^{\rho}\phi-2\lambda F^{\rho\mu}\nabla_{\rho}\chi,
\end{aligned} \label{eq:max}
\ee
and the linearized Klein-Gordon equation $\delta \hat S = 0$, where 
\be
\begin{aligned}
\delta \hat S & =(3\lambda^{2}-1)\left(\nabla_{\mu}\phi\nabla^{\mu}h_{\nu}^{\;\;\nu}-2\nabla_{\mu}\phi\nabla_{\nu}h^{\mu\nu}-2h^{\mu\nu}\nabla_{\mu}\nabla_{\nu}\phi+2\lambda h^{\mu\nu}\nabla_{\mu}\phi\nabla_{\nu}\phi\right)\\
&\qquad +2(3\lambda^{2}-1)\left(\Box\chi-2\lambda\nabla_{\mu}\phi\nabla^{\mu}\chi\right)\\
 & \qquad+\lambda\nabla_{\mu}\nabla_{\nu}h^{\mu\nu}-\lambda\Box h_{\mu}^{\;\;\mu}-\lambda h^{\mu\nu}R_{\mu\nu}-\lambda F^{\mu\nu}f_{\mu\nu}+\lambda F_{\mu\rho}F_{\nu}^{\;\;\rho}h^{\mu\nu}.
\end{aligned}\label{eq:kg}
\ee

For the $\ell \geq 2$ polar modes, we write the perturbations as~\cite{Dolan,Pani,Cardoso}
\be 
\begin{aligned}
h_{\mu\nu}&=\left(\begin{array}{cccc}
\left(\frac{1+\lambda^{2}}{1-\lambda^2}\right)^{2}\frac{r_{0}^{2}}{z^{2}}H_{0}(z) & H_{1}(z) & 0 & 0\\
H_{1}(z) & \left(\frac{1+\lambda^{2}}{1-\lambda^2}\right)^{2}\frac{r_{0}^{2}}{z^{2}}H_{2}(z) & 0 & 0\\
0 & 0 & r_{0}^{2}K(z) & 0\\
0 & 0 & 0 & r_{0}^{2}K(z)\sin^{2}\theta
\end{array}\right)\times e^{i\omega t}P_{\ell}(\cos\theta) \\
a_\mu &= \left(0,0,0,u_{4}(z)\right)\times e^{i\omega t} \sin\theta \tfrac{{\rm d}}{{\rm d}\theta}P_{\ell}(\cos\theta) \\
\chi &= u_{0}(z)e^{i\omega t}P_{\ell}(\cos\theta),
\end{aligned}\label{eq:polar}
\ee
where $P_\ell$ is the Legendre polynomial of degree $\ell$. For the $\ell \geq 2$ axial modes, we write
\be 
\begin{aligned}
h_{\mu\nu}&=\left(\begin{array}{cccc}
0 & 0 & 0 & h_{0}(z)\\
0 & 0 & 0 & h_{1}(z)\\
0 & 0 & 0 & 0\\
h_{0}(z) & h_{1}(z) & 0 & 0
\end{array}\right)\times e^{i\omega t}\sin\theta\tfrac{{\rm d}}{{\rm d}\theta}P_{\ell}(\cos\theta) \\
a_{\mu}&=\left(u_{1}(z)P_{\ell}(\cos\theta),u_{2}(z)P_{\ell}(\cos\theta),u_3(z) \tfrac{{\rm d}}{{\rm d}\theta}P_\ell (\cos\theta),0\right)\times e^{i\omega t}  \\
\chi &= 0.
\end{aligned}\label{eq:axial}
\ee
The perturbations above are in so-called Regge-Wheeler-Zerilli gauge \cite{ReggeWheeler,Zerilli,ZerilliRN}.
A ${\rm U}(1)$ gauge choice allows us to set $u_3(z)$ to zero, which we do hereafter, thereby completely fixing the gauge.
In the Einstein-Maxwell-dilaton equations of motion, the polar and axial modes decouple.

We note that if we were to redefine dilaton fields via $\phi \rightarrow \lambda \phi$, $\chi\rightarrow \lambda\chi$, and rescale the Klein-Gordon equation $\delta\hat S \rightarrow \lambda \delta\hat S$, then our equations of motion in Eqs.~\eqref{eq:ein}, \eqref{eq:max}, and \eqref{eq:kg}, as well as our background in \Eq{eq:metstring}, would all be even in $\lambda$.
As a result, the spectra of perturbations described by the towers of $g$ couplings in \Eq{eq:gs} are also all even in $\lambda$, and we therefore take $\lambda>0$ without loss of generality henceforth.

\section{Axial modes}\label{sec:axial}
Let us first consider the $\ell\geq 2$ axial modes: those of odd parity for the graviton, even parity for the gauge field, and with vanishing dilaton perturbation.
Taking the axial form of the perturbations in \Eq{eq:axial}, we rewrite the perturbative Einstein, Maxwell, and Klein-Gordon equations in Eqs.~\eqref{eq:ein}, \eqref{eq:max}, and \eqref{eq:kg} in terms $h_0(z)$, $h_1(z)$, $u_1(z)$, and $u_2(z)$.
We wish to find the master variables and their associated effective potentials.
We find that the equation $\delta \hat E_\theta^{\;\;\varphi} = 0$ requires
\be
i\omega h_{0}(z)=h_{1}'(z)-\frac{2\lambda^{2}}{1-\lambda^2}\frac{h_{1}(z)}{z}, \label{eq:h0}
\ee
which we set henceforth.
Having done so, we now find that there are five nonvanishing, a priori independent, components of the equations of motion, $\delta \hat M^t$, $\delta \hat M^z$, $\delta \hat M^\theta$, $\delta \hat E_t^{\;\;\varphi}$, and $\delta \hat E_z^{\;\;\varphi}$.
We will find it convenient to relabel these five equations as follows:
\be
\begin{aligned}
\alpha_{0} & =-\frac{(1+\lambda^{2})^{8}m^{4}}{(1-\lambda^2)^{2}z^{2}P_{\ell}(\cos\theta)}e^{-i\omega t}\delta \hat M^{t}\\
\alpha_{1} & =-\frac{(1+\lambda^{2})^{8}m^{4}}{(1-\lambda^2)^{2}z^{2}P_{\ell}(\cos\theta)}e^{-i\omega t}\delta \hat M^{z}\\
\alpha_{2} & =-\frac{(1+\lambda^{2})^{6}m^{4}}{(1-\lambda^2)z\frac{{\rm d}}{{\rm d}\theta}P_{\ell}(\cos\theta)}e^{-i\omega t}\delta\hat M^{\theta}\\
\alpha_{3} & =-\frac{(1-\lambda^2)(1+\lambda^{2})^{6}m^{4}\omega z\sin\theta}{\frac{{\rm d}}{{\rm d}\theta}P_{\ell}(\cos\theta)}e^{-i\omega t}\delta\hat E_{t}^{\;\;\varphi}\\
\alpha_{4} & =\frac{2(1+\lambda^{2})^{6}m^{4}\sin\theta}{\frac{{\rm d}}{{\rm d}\theta}P_{\ell}(\cos\theta)}e^{-i\omega t}\delta\hat E_{z}^{\;\;\varphi}.
\end{aligned} \label{eq:alphas}
\ee
By construction, all of the $\alpha_i$ are purely functions of $z$. Requiring $\alpha_2 = 0$ implies
\be 
i\omega u_{1}(z)= u_{2}'(z)-\frac{2\lambda^{2}}{1-\lambda^2}\frac{u_{2}(z)}{z},
\ee
which we also set henceforth.
We then find that the following relations among the $\alpha_i$ are identically satisfied:
\be 
\begin{aligned}
i\omega\alpha_{0}+\frac{{\rm d}\alpha_{1}}{{\rm d}z}-\frac{2\lambda^{2}}{1-\lambda^2}\frac{\alpha_{1}}{z} & =0\\
2i\alpha_{3}+(1-\lambda^2)z\frac{{\rm d}\alpha_{4}}{{\rm d}z}-2\lambda^{2}\alpha_{4} & =0.
\end{aligned}\label{eq:alpharedundancies}
\ee
We use the above relations to eliminate the equations of motion $\alpha_{0}=0$ and $\alpha_{3}=0$ as redundant.
Our only remaining equations of motion are thus $\alpha_{1}=0$ and $\alpha_{4}=0$, in terms of $u_{2}(z)$ and $h_{1}(z)$, giving a set of coupled second-order differential equations. We can remove the first-order terms $h_1'$ and $u_2'$ by defining rescaled functions:
\be
\begin{aligned}
\bar{h}_{1}(z) & = z^{-\frac{\lambda^{2}}{1-\lambda^2}}h_{1}(z)\\
\bar{u}_{2}(z) & = z^{-\frac{\lambda^{2}}{1-\lambda^2}}u_{2}(z).
\end{aligned} 
\ee
The resulting system of equations for $\bar h_1$ and $\bar u_2$ can be written in elegant matrix form:
\be 
\left(\frac{{\rm d}^{2}}{{\rm d}z^{2}}+\omega^{2}\right)\vec{\mathbf{x}}=\frac{1}{z^{2}}\mathbf{A}\vec{\mathbf{x}},\label{eq:matrixeq}
\ee
where we have defined a constant matrix,
\be 
\mathbf{A}=\left(\begin{array}{cc}
\frac{2-\lambda^{2}+\ell(\ell+1)(1+\lambda^{2})^{2}}{(1-\lambda^2)^2} & -2\sqrt{2}\left(\frac{1+\lambda^{2}}{1-\lambda^2}\right)^2\\
-\sqrt{2}\frac{\ell(\ell+1)(1+\lambda^{2})}{(1-\lambda^2)^2} & \frac{\ell(\ell+1)+\lambda^{2}[2\ell(\ell+1)-1]+\lambda^{4}(\ell^{2}+\ell+2)}{(1-\lambda^2)^2}
\end{array}\right),
\ee
and a vector describing our degrees of freedom,
\be 
\vec{\mathbf{x}}=\left(\begin{array}{c}
\bar{h}_{1}(z)\\
m\sqrt{1+\lambda^{2}}\bar{u}_{2}(z)
\end{array}\right).
\ee
Diagonalizing $\bf A$, we find the eigenvalues $g^{\rm ax}_\pm$ as given in \Eq{eq:gs}. Using the eigenvectors to define new field variables that mix the photon and graviton modes,
\be
y_\pm (z) = \ell(\ell+1)\bar h_1(z) + \sqrt{\frac{1{+}\lambda^2}{2}}\left[1-\lambda^2 \mp \sqrt{(\lambda^2 + 2\ell+1)^2 + 4\lambda^2(\ell+1)(\ell-1)}\right]m\,\bar u_2(z),\label{eq:ypm}
\ee
we find that the resulting equations of motion drastically simplify:
\be
y_{\pm}''(z)+\omega^{2}y_{\pm}(z)-\frac{g^{\rm ax}_{\pm}}{z^{2}}y_{\pm}(z)=0.
\ee
Here, $g^{\rm ax}_\pm /z^2$ is the effective potential for the mixed photon/graviton wave function of our ${\rm AdS}_2 \times S^2$ geometry.
The $\pm$ choice reflects the fact that the axial sector contains two distinct towers of propagating modes, indexed by angular momentum $\ell$.

\section{Polar modes}\label{sec:polar}
Having identified the axial propagating modes for our ${\rm AdS}_2 \times S^2$ solution, we now turn to the polar case for $\ell\geq 2$, where the perturbations are of even parity for the graviton and dilaton and odd parity for the gauge field.
We will find that the equations of motion are more complicated in the case of polar modes.
As our starting point, we take the polar form of the perturbations in \Eq{eq:polar} and rewrite the perturbed Einstein, Maxwell, and Klein-Gordon equations given in Eqs.~\eqref{eq:ein}, \eqref{eq:max}, and \eqref{eq:kg} in terms $H_0(z)$, $H_1(z)$, $H_2(z)$, $K(z)$, $u_0(z)$, and $u_4(z)$.

Requiring $\delta \hat E_\theta^{\;\;\theta} - \delta \hat E_\varphi^{\;\;\varphi} = 0$  fixes
\be
H_{2}(z)=H_{0}(z)+4\lambda u_{0}(z). 
\ee
Meanwhile, $\delta \hat E_t^{\;\;z} = 0$ implies
\be 
H_{1}(z)=\frac{2i\omega r_{0}^{2}}{\ell(\ell+1)z}\left\{ K(z)+zK'(z)+\frac{\lambda^{2}}{1-\lambda^2}\left[H_{0}(z)+4\lambda u_{0}(z)\right]-2\lambda zu_{0}'(z)\right\} .
\ee
Of the remaining a priori independent components of the equations of motion, we find there are seven that do not already vanish.
In analogy with \Eq{eq:alphas}, we will write these seven equations as:
\be
\begin{aligned}\beta_{0} & =-\frac{m^{4}(1+\lambda^{2})^{6}\sin\theta}{z^{2}\frac{{\rm d}}{{\rm d}\theta}P_{\ell}(\cos\theta)}e^{-i\omega t}\delta\hat M^{\varphi}\\
\beta_{1} & =-\frac{m^{2}(1+\lambda^{2})^{6}}{z^{2}P_{\ell}(\cos\theta)}e^{-i\omega t}(\delta\hat E_{t}^{\;\;t}-\delta\hat E_{z}^{\;\;z})\\
\beta_{2} & =\frac{m^{2}(1+\lambda^{2})^{6}}{z^{2}P_{\ell}(\cos\theta)}e^{-i\omega t}(\delta\hat E_{t}^{\;\;t}+\delta \hat E_{z}^{\;\;z})\\
\beta_{3} & =\frac{2i\ell(\ell+1)m^{3}(1+\lambda^{2})^{4}}{\omega\frac{{\rm d}}{{\rm d}\theta}P_{\ell}(\cos\theta)}e^{-i\omega t}\delta\hat E_{t}^{\;\;\theta}\\
\beta_{4} & =\frac{2\ell(\ell+1)m^{3}z(1+\lambda^{2})^{4}}{\frac{{\rm d}}{{\rm d}\theta}P_{\ell}(\cos\theta)}e^{-i\omega t}\delta \hat E_{z}^{\;\;\theta}\\
\beta_{5} & =\frac{2\ell(\ell+1)m^{3}(1+\lambda^{2})^{6}}{P_{\ell}(\cos\theta)}e^{-i\omega t}\delta \hat E_{\theta}^{\;\;\theta}.\\
\beta_{6} & =\frac{\ell(\ell+1)m^{3}(1+\lambda^{2})^{6}}{P_{\ell}(\cos\theta)}e^{-i\omega t}\delta \hat S.
\end{aligned}\label{eq:betas}
\ee
By construction, the $\beta_i$ are functions solely of $z$. 
We have four remaining unfixed functions---$H_0$, $K$, $u_0$, and $u_4$---which we rescale as
\be 
\begin{aligned}\bar{H}_{0}(z) & =z^{-\frac{\lambda^{2}}{1-\lambda^2}}H_{0}(z)\\
\bar{K}(z) & =z^{-\frac{\lambda^{2}}{1-\lambda^2}}K(z)\\
\bar{u}_{0}(z) & =z^{-\frac{\lambda^{2}}{1-\lambda^2}}u_{0}(z)\\
\bar{u}_{4}(z) & =z^{-\frac{\lambda^{2}}{1-\lambda^2}}u_{4}(z)
\end{aligned}\label{eq:rescale}
\ee
and similarly define $\bar{\beta}_{i}=z^{-\frac{\lambda^{2}}{1-\lambda^2}}\beta_{i}$.
The seven equations of motion~\eqref{eq:betas} are not overdetermined, since there are three identically satisfied relations among the $\bar{\beta}_{i}$:
\be 
\begin{aligned}z^{2}m\bar{\beta}_{1}-(1+\lambda^{2})^{2}\bar{\beta}_{3}+\frac{\bar{\beta}_{5}}{\ell(\ell+1)} & =0\\
2z^{2}m(1-2\lambda^{2})\bar{\beta}_{1}+(1+\lambda^{2})^{2}\left[\lambda^{2}\bar{\beta}_{3}-(1-\lambda^2)\bar{\beta}_{4}\right]-\frac{2\lambda\bar{\beta}_{6}}{\ell(\ell+1)}\\
+(1-\lambda^2)\left[2mz^{3}\frac{{\rm d}\bar{\beta}_{1}}{{\rm d}z}-(1+\lambda^{2})^{2}z\frac{{\rm d}\bar{\beta}_{3}}{{\rm d}z}\right] & =0\\
2\sqrt{2}\sqrt{1+\lambda^{2}}\bar{\beta}_{0}-m\bar{\beta}_{2}-\frac{1-\lambda^2}{\ell(\ell+1)}\left[\frac{\bar{\beta}_{4}}{z^{2}}-\frac{1-\lambda^2}{z}\frac{{\rm d}\bar{\beta}_{4}}{{\rm d}z}+(1-\lambda^2)\omega^{2}\bar{\beta}_{3}\right] & =0.
\end{aligned}\label{eq:betaredundancies}
\ee
We choose subsequently to drop $\bar\beta_2$, $\bar\beta_5$, and $\bar\beta_6$ as redundant.
For the four remaining equations of motion $\bar\beta_{0,1,3,4}$, let us redefine as follows:
\be
\begin{aligned}
\hat\beta_1 &= \frac{\ell(\ell+1)}{4\lambda^4 (1-\lambda^2)^2 z^3} \left[2 m z \bar\beta_1 - (1+\lambda^2)^2 \frac{\bar\beta_3}{z} - 2\lambda^2(1-\lambda^2)\frac{\bar\beta_4}{\ell(\ell+1)z} \right]\\
\hat\beta_2 &= -\frac{1}{4\lambda (1-\lambda^2)}\left[2mz \bar\beta_1 - (1+\lambda^2)^2 \frac{\bar\beta_3}{z} - 4\lambda^2(1-\lambda^2)\frac{\bar\beta_4}{\ell(\ell+1)z}\right] \\
\hat\beta_3 &= \bar\beta_0\\
\hat\beta_4 &= 2\sqrt{2}\ell(\ell+1)\lambda^2\sqrt{1+\lambda^2}z^2 \bar\beta_0 - 2\lambda^4(1-\lambda^2)^3 z^2 \frac{{\rm d}}{{\rm d}z}(z^3 \hat\beta_1)  + 2\lambda^4(1-\lambda^2)^2 z^4 \hat \beta_1 \\& \qquad -\lambda^2 (1-\lambda^2)^2 \omega^2 z^2 \bar\beta_3 + \frac{1}{2}(1-\lambda^2)[\lambda^2(\ell-1)+\ell+1][\lambda^2(\ell+2)+\ell]\bar\beta_4.
\end{aligned} \label{eq:betahats}
\ee
By construction, we have $\bar\beta_0 =  \bar\beta_1 =  \bar \beta_3 =  \bar\beta_4 = 0$ if and only if $\hat\beta_1 = \hat\beta_2 = \hat\beta_3 = \hat\beta_4 = 0$.
Remarkably, the equations $\hat\beta_3 = 0$ and $\hat\beta_4 = 0$ decouple from the dilaton, depending only on the photon and graviton modes $\bar u_4$ and $\bar K$. 
We conjecture that this miracle can be traced to ${\rm SL}(2)$ symmetry: the ${\rm AdS}_2\times S^2$ metric and gauge field are themselves invariant under the ${\rm SL}(2)$ transformations comprised of time translation, dilations in $z$, and special conformal transformations, but the dilaton background $\phi$ is not.
We therefore conjecture that it is the ${\rm SL}(2)$ invariance of the photon/graviton sector of the theory that is responsible for the clean separation of these modes.
The requirements $\hat \beta_{3,4} = 0$ give two coupled second-order differential equations that can be written in matrix form as in \Eq{eq:matrixeq}, but in terms of a new constant matrix,
\be
\mathbf{A}=\left(\begin{array}{cc}
\frac{\ell^{2}+\ell+2+\lambda^{2}(2\ell^{2}+2\ell+5)+\lambda^{4}(\ell^{2}+\ell+2)}{(1-\lambda^2)^2} & 2\sqrt{2}\ell(\ell+1)\left(\frac{1+\lambda^2}{1-\lambda^2}\right)^2\\
\sqrt{2}\left(\frac{1+\lambda^2}{1-\lambda^2}\right)^2 & \frac{[(1+\lambda^{2})\ell+1][(\ell+1)\lambda^{2}+\ell]}{(1-\lambda^2)^2}
\end{array}\right), \label{eq:A2}
\ee
and a new vector describing our photon and graviton degrees of freedom,
\be 
\vec{\mathbf{x}}=\left(\begin{array}{c}
\bar{K}(z)\\
\bar{u}_{4}(z)/m\sqrt{1+\lambda^{2}}
\end{array}\right).
\ee
Diagonalizing $\bf A$ in \Eq{eq:A2}, we find the eigenvalues are $g_\pm^{\rm pol}$ as given in \Eq{eq:gs}.
Redefining our degrees of freedom using the eigenvectors, we write the mixed photon/graviton modes,
\be
\begin{aligned}
w_+(z) &= \frac{1}{\sqrt{2}}\bar K(z) + \frac{\ell}{m\sqrt{1+\lambda^2}} \bar u_4(z) \\
w_-(z) &= -\frac{1}{\sqrt{2}}\bar K(z) + \frac{\ell + 1}{m\sqrt{1+\lambda^2}} \bar u_4(z).
\end{aligned} \label{eq:wpm}
\ee
In terms of these mixed degrees of freedom, our equations of motion for these two modes simplify substantially:
\be 
w''_\pm(z) + \omega^2 w_\pm(z)  - \frac{g_\pm^{\rm pol}}{z^2} w_\pm(z) = 0.\label{eq:eomwpm}
\ee

What about the remaining two equations of motion, $\hat \beta_1 = \hat\beta_2 = 0$?
Explicit evaluation reveals these both to be first-order---albeit lengthy---expressions in terms of $\bar H_0$,  $\bar u_0$, $\bar K$, and $\bar u_4$, so we should expect a single final propagating mode.
To isolate this final mode, we will find it useful to first exchange $\bar H_0(z)$ and $\bar u_0(z)$ for two new functions $j(z)$ and $w_0(z)$ defined via
\be 
\begin{aligned} 
& \frac{1+\lambda^{2}}{1-\lambda^2}\left[\lambda^{2}(\ell-1)+\ell+1\right]\left[\lambda^{2}(\ell+2)+\ell\right]\left[\bar{H}_{0}(z)-z^{2}w_{0}(z)+4\lambda j(z)\right]\\
 & =2(1-\lambda^2)z\left\{ \bar{K}'(z)-\omega^{2}z\bar{K}(z)-\frac{2\sqrt{2}\lambda^{2}}{(1+\lambda^{2})^{3/2}m}\left[\bar{u}_{4}'(z)-\omega^{2}z\bar{u}_{4}(z)\right]\right\} \\
 & \qquad+\left[\ell(\ell+1)(1+\lambda^{2})^{2}+2(1+\lambda^{2}-\lambda^{4})\right]\left[\bar{K}(z)+\frac{2\sqrt{2}}{(1+\lambda^{2})^{3/2}m}\bar{u}_{4}(z)\right]\\
 & \qquad-\frac{4\sqrt{2}}{\sqrt{1+\lambda^{2}}m}\bar{u}_{4}(z)
\end{aligned}\label{eq:qz}
\ee
and 
\be 
\begin{aligned} 
& (1+\lambda^{2})\left[\lambda^{2}(\ell-1)+\ell+1\right]\left[\lambda^{2}(\ell+2)+\ell\right]\left[\bar{u}_{0}(z)-j(z)\right]\\
 & =\lambda(1-\lambda^2)z\left[\bar{K}'(z)-\frac{2\sqrt{2}\lambda^{2}}{(1+\lambda^{2})^{3/2}m}\bar{u}_{4}'(z)\right]\\
 & \qquad+\frac{\lambda\left[\ell(\ell+1)(1+\lambda^{2})^{2}+2(1+\lambda^{2}-\lambda^{4})\right]}{2}\left[\bar{K}(z)+\frac{2\sqrt{2}}{(1+\lambda^{2})^{3/2}m}\bar{u}_{4}(z)\right]\\
 & \qquad-\frac{2\sqrt{2}\lambda}{\sqrt{1+\lambda^{2}}m}\bar{u}_{4}(z).
\end{aligned}\label{eq:w0z}
\ee
We also exchange $\hat\beta_1$ and $\hat \beta_2$ for
\be
\begin{aligned}
\check\beta_1 &=  \hat\beta_1 - \frac{2\omega^2 \left[2\sqrt{2}\lambda^2 \hat\beta_3 - \frac{1}{\ell(\ell+1)\sqrt{1+\lambda^2}z^2}\hat\beta_4 \right]}{(1+\lambda^2)^{5/2}[\lambda^2(\ell-1)+\ell+1][\lambda^2(\ell+2)+\ell]} \\
\check\beta_2 &= \hat\beta_2 + \frac{\lambda\sqrt{1+\lambda^2}z \left[2\sqrt{2}\lambda^2 \hat\beta_3 - \frac{1}{\ell(\ell+1)\sqrt{1+\lambda^2}z^2}\hat\beta_4 \right]}{(1-\lambda^2)^2[\lambda^2(\ell-1)+\ell+1][\lambda^2(\ell+2)+\ell]}.
\end{aligned}\label{eq:betachecks}
\ee
Given that $\hat\beta_3 =0$ and $\hat\beta_4 = 0$ are satisfied (which defines the two modes $w_\pm$ that we have already found), satisfying $\check\beta_1 = \check\beta_2 = 0$ is equivalent to satisfying $\hat\beta_1 = \hat\beta_2 = 0$.
In terms of our new variables $w_0(z)$ and $j(z)$, $\check\beta_1 = 0$ and $\check\beta_2 = 0$ define the following pair of coupled first-order equations:
\be
\begin{aligned} & \omega^{2}j'(z)+\frac{\left[\ell(\ell+1)(1+\lambda^{2})^{2}+2\lambda^{4}\right]\omega^{2}}{2\lambda^{2}(1-\lambda^2)z}j(z)\\
 & -\left\{ \frac{\ell(\ell+1)(1+\lambda^{2})^{2}\left[\lambda^{2}(\ell-1)+\ell+1\right]\left[\lambda^{2}(\ell+2)+\ell\right]}{8\lambda^{3}(1-\lambda^2)^{3}z}+\frac{\lambda\omega^{2}z}{2(1-\lambda^2)}\right\} w_{0}(z) =0\label{eq:eomfinal1}
\end{aligned}
\ee
and 
\be 
w_{0}'(z)-\frac{\ell(\ell+1)(1+\lambda^{2})^{2}+2\lambda^{2}}{2\lambda^{2}(1-\lambda^2)z}w_{0}(z)+\frac{2(1-\lambda^2)}{\lambda z}\omega^{2}j(z)=0.\label{eq:eomfinal2}
\ee
We see that \Eq{eq:eomfinal2} is algebraic for $j(z)$, so solving it and
inputting the solution  into \Eq{eq:eomfinal1}, we at last find the equation of motion for $w_0(z)$,
\be 
w_0''(z) + \omega^2 w_0(z) - \frac{g^{\rm pol}_0}{z^2} w_0(z) = 0,\label{eq:eomw0}
\ee
where the coupling $g_0^{\rm pol}$ is given in \Eq{eq:gs}.
Together, Eqs.~\eqref{eq:eomwpm} and \eqref{eq:eomw0} are the wave equations for polar modes of the coupled graviton/photon/dilaton system about our ${\rm AdS}_2\times S^2$ background, comprising three distinct towers indexed by $\ell$.

\section{Dipole modes}\label{sec:l1}

Finding the modes in the $\ell=1$ case proceeds in much the same way as the $\ell \geq 2$ cases considered in Secs.~\ref{sec:axial} and \ref{sec:polar}.
The main difference will be that some of the previous equations end up vanishing identically, indicating that a subset of $\ell \geq 2$ degrees of freedom found above become pure gauge for $\ell = 1$.
Among the five towers of physical degrees of freedom for each $\ell \geq 2$ (two axial and three polar), we will find three (one axial and two polar)  at $\ell = 1$.

In general, given a diffeomorphism $x^\mu \rightarrow x^\mu + \xi^\mu$, we have the pure gauge perturbations
\be
\begin{aligned}
\delta h_{\mu\nu} &= {\cal L}_\xi g_{\mu\nu} &&\!\!\!\!\!\! = \nabla_\mu \xi_\nu + \nabla_\nu \xi_\mu \\
\delta a_\mu &= {\cal L}_\xi A_\mu &&\!\!\!\!\!\!  = \xi^\nu \nabla_\nu A_\mu + A_\nu \nabla_\mu \xi^\nu \\
\delta\chi &= {\cal L}_\xi \phi &&\!\!\!\!\!\!  = \xi^\mu \nabla_\mu \phi,
\end{aligned} 
\ee
where $g_{\mu\nu}$, $A_\mu$, and $\phi$ describe the background solution~\eqref{eq:metstring} in string frame.
The Regge-Wheeler-Zerilli ansatz of Eqs.~\eqref{eq:polar} and \eqref{eq:axial} does not completely fix the gauge for $\ell=1$ modes.
As we will see, this additional gauge freedom will eat some of the erstwhile physical modes for $\ell = 1$.

\subsection{Axial vector}\label{sec:axialvector}

For the axial vector perturbation, starting with the ansatz in \Eq{eq:axial}, we find that $\delta \hat E_\theta^{\;\;\varphi}$ now vanishes identically because it is proportional to $2\cot\theta \frac{{\rm d}}{{\rm d}\theta} P_\ell(\cos\theta) + \ell(\ell+1) P_\ell(\cos\theta)$, which is zero for $\ell=1$ but not for general $\ell$.
We therefore do not have $h_0$ determined in terms of $h_1$ as in \Eq{eq:h0}.  
On the other hand, for $\ell=1$ the ansatz in \Eq{eq:axial} is not fully gauge-fixed because there is residual gauge freedom, within the axial ansatz~\eqref{eq:axial}, generated by $\xi=\Xi(t,z)\partial_\phi$. We may use this gauge freedom to set $u_1=0$.
For propagating ($\omega\neq 0$) modes, this choice completely fixes the gauge.

From $\delta \hat M^{\theta}=0$ we now find an expression for $h_{0}(z)$ in terms of $h_1$ and $u_2$,
\be 
i\omega h_{0}(z)=h_{1}'(z)-\frac{2\lambda^{2}}{1-\lambda^2}\frac{h_{1}(z)}{z}-\frac{m(1+\lambda^{2})^{3/2}}{\sqrt{2}}\left[u_{2}'(z)-\frac{2\lambda^{2}}{1-\lambda^2}\frac{u_{2}(z)}{z}\right],
\ee
which we set henceforth.
Then the remaining nonvanishing equations of motion are $\delta \hat M^t$, $\delta\hat M^z$, $\delta \hat E_t^{\;\;\varphi}$, and $\delta\hat E_z^{\;\;\varphi}$. Defined in terms of the $\alpha_i$ in \Eq{eq:alphas} (with $\delta\hat M^\theta$ vanishing), two of these four equations are redundant precisely as in \Eq{eq:alpharedundancies}, so it will suffice to enforce the two equations $\delta\hat M^z =0$ and $\delta \hat E_z^{\;\;\varphi}=0$. We find that $\delta\hat M^z=0$ fixes $h_{1}(z)$:
\be 
h_{1}(z)=\frac{m}{2\sqrt{2}\sqrt{1+\lambda^{2}}} \left[2(1+\lambda^{2})^{2}-\omega^{2}z^{2}(1-\lambda^2)^{2}\right]u_{2}(z).\label{eq:h1set}
\ee 
 In terms of $z^2 \bar u_2(z) = z^{2-\frac{\lambda^2}{1-\lambda^2}}u_2(z)$, we find an equation of motion of the form~\eqref{eq:conformalQM},
\be 
\frac{{\rm d}^2}{{\rm d}z^2} [z^2 \bar u_2(z)]+\omega^{2} [z^2 \bar u_2(z)]-\frac{6+7\lambda^{2}+4\lambda^{4}}{(1-\lambda^2)^{2}z^{2}}[z^2 \bar u_2(z)]=0.\label{eq:u2eoml1}
\ee
We note that the coefficient of the $1/z^{2}$ potential is precisely $g_+^{\rm ax}$ in Eq.~\eqref{eq:gs}, evaluated at $\ell=1$. Moreover, the $\ell \geq 2$ degrees of freedom defined in \Eq{eq:ypm} can for $\ell=1$ be written in terms of $\bar u_2(z)$ alone, by virtue of \Eq{eq:h1set}. Specifically, one finds that, for $\ell=1$, $y_+(z)$ from \Eq{eq:ypm} is proportional to $z^2 \bar u_2(z)$ and hence is the degree of freedom in Eq.~\eqref{eq:u2eoml1}.
Meanwhile, $y_-$ is no longer independent; for $\ell=1$ it is defined strictly in terms of $y_+$ (or, equivalently, $\bar u_2$).
There is thus only one physical propagating axial vector mode. In the Weyl gauge where $a_t=0$, since $y_+(z) \propto z^2 \bar u_2(z)$ by \Eq{eq:h1set}, we have found that this axial mode is associated with the $z$-component of the gauge potential perturbation.

\subsection{Polar vectors}

We now turn to the polar $\ell = 1$ case, where the graviton and dilaton perturbations are parity-even and the photon perturbation is parity-odd as in \Eq{eq:polar}.
The main difference here between $\ell = 1$ and $\ell \geq 2$ is that, for the vector perturbation, setting $\delta\hat E_{\theta}^{\;\;\theta}-\delta\hat E_{\varphi}^{\;\;\varphi}=0$ no longer requires $H_{2}(z)=H_{0}(z)+4\lambda u_{0}(z)$, so we do not impose this. 
The reason, as in the axial case in \Sec{sec:axialvector}, is that this equation of motion satisfies $\delta\hat E_{\theta}^{\;\;\theta}-\delta\hat E_{\varphi}^{\;\;\varphi}\propto 2\cot\theta\frac{{\rm d}}{{\rm d}\theta}P_{\ell}(\cos\theta)+ \ell(\ell+1)P_{\ell}(\cos\theta)$, which vanishes identically for the $\ell=1$ case.
This is balanced by the fact that for $\ell=1$ the ansatz in \Eq{eq:polar} is not fully gauge-fixed. Indeed, for  $\ell=1$ there is residual gauge freedom, within the polar ansatz~\eqref{eq:polar}, generated by
\be
\xi=\Xi^t(t,z)\cos\theta \, \partial_t 
+\Xi^z(t,z)\cos\theta \, \partial_z
-\Xi(t,z)\sin\theta \, \partial_\theta,
\ee
with $\Xi^t=\left(\frac{1-\lambda^2}{1+\lambda^2}\right)^2 z^2 \, \partial_t \Xi$ and $\Xi^z=-\left(\frac{1-\lambda^2}{1+\lambda^2}\right)^2 z^2 \, \partial_z \Xi$. We may use this gauge freedom to set $K=0$, thereby completely fixing the gauge.

Fixing $\delta\hat E_t^{\;\;z} = 0$ implies
\be
 H_1(z) = \frac{i\omega r_0^2}{z}\left[\frac{\lambda^2}{1-\lambda^2}H_2(z) - 2 \lambda z u_0'(z) \right],
\ee
which we set henceforth.
There are seven a priori independent equations of motion that do not identically vanish, which we can write as $\beta_i$ as in \Eq{eq:betas}.
We rescale the functions as in \Eq{eq:rescale}, as well as $\bar H_2(z) = z^{-\frac{\lambda^2}{1-\lambda^2}} H_2(z)$.
Defining $\bar \beta_i = z^{-\frac{\lambda^2}{1-\lambda^2}} \beta_i$ as before, we find that the three relations in \Eq{eq:betaredundancies} are still satisfied for $\ell=1$, leaving us with four independent equations and four undetermined functions, which we package in terms of the $\hat\beta_i$ as in \Eq{eq:betahats}.

From $\hat\beta_4 = 0$, we fix $\bar H_2(z)$:
\be
\bar H_2(z) =  \bar H_0(z) + 4\lambda \bar u_0(z) -\frac{4\sqrt{2}}{m\sqrt{1+\lambda^2}}\, \bar u_4(z).
\ee
From $\hat\beta_3 = 0$ we then find an equation of motion for $\bar u_4(z)$ alone:
\be
\bar u_4''(z) + \omega^2 \bar u_4(z) - \frac{(2+3\lambda^2)(3+2\lambda^2)}{(1-\lambda^2)^2 z^2} \bar u_4(z) = 0.
\ee
The coefficient of the $1/z^2$ potential is precisely $g_+^{\rm pol}$ in \Eq{eq:gs}, evaluated at $\ell = 1$.
In our $\bar K = 0$ gauge, the two degrees of freedom $w_\pm$ in \Eq{eq:wpm} have merged into one given by $\bar u_4$.
For the final, dilatonic mode, we define $j(z)$ and $w_0(z)$ as in Eqs.~\eqref{eq:qz} and \eqref{eq:w0z}, along with $\check \beta_{1,2}$ as in \Eq{eq:betachecks}, all for $\ell = 1$, and running through the logic as in \Sec{sec:polar}, we find the final equation of motion,
\be
w_0''(z) + \omega^2 w_0(z) - \frac{(1+2\lambda^2)(2+\lambda^2)}{(1-\lambda^2)^2 z^2} w_0(z) = 0,
\ee
where the coefficient of the potential is $g_0^{\rm pol}$ in \Eq{eq:gs}, evaluated for $\ell = 1$.
There are thus two propagating polar vector modes in our charged, dilatonic ${\rm AdS}_2 \times S^2$ background.

\section{Monopole mode}\label{sec:l0}

The spherically symmetric $\ell=0$ perturbations are of polar type. 
There exist several non-propagating $\ell=0$ perturbations that are solutions to the linearized Eqs.~\eqref{eq:ein}, \eqref{eq:max}, and \eqref{eq:kg}. 
The simplest such solution is given by a parameter variation $r_0\to r_0+\delta r_0$ of our ${\rm AdS}_2 \times S^2$ background in \Eq{eq:metstring}. 
Another static solution is the perturbation that is obtained from the leading-order correction to the near-horizon scaling limit \eqref{eq:scaling} that produces our ${\rm AdS}_2 \times S^2$ from extreme GHS. 
Such solutions, the so-called anabasis perturbations, were recently discussed in depth for Bertotti-Robinson in Ref.~\cite{Hadar:2020kry}. 
On the other hand, in the Einstein-Maxwell-dilaton theory in this paper, the presence of the dilaton implies the existence of a single $\ell=0$ propagating wave degree of freedom as well. 
Let us find the corresponding wave equation.

For $\ell=0$ the ansatz in \Eq{eq:polar} is spherically symmetric but it is neither gauge-fixed nor general enough with respect to the Maxwell field perturbation. For the latter we must allow\footnote{For a magnetic monopole there is no globally well defined vector potential $A$, so in this section we work directly with the perturbation for the Maxwell field $F$.}  
\be
\delta F=f_{\theta\varphi}(t,z) \sin \theta \,{\rm d}\theta \wedge {\rm d}\varphi\,,
\ee
and in a mode expansion set $f_{\theta\varphi}=v(z)e^{i\omega t}$.
Together with $h_{\mu\nu}$ and $\chi$ from the $\ell=0$ version of \Eq{eq:polar}, this perturbation gives the most general spherically symmetric ansatz of polar type. To fix the gauge we note that an $\ell=0$ diffeomorphism generated by the vector field $\xi=\xi^t(t,z)\partial_t+\xi^z(t,z)\partial_z$ shifts only $h_{tt}$, $h_{tz}$, $h_{zz}$, and $\chi$, leaving $h_{\theta\theta}$ and $f_{\theta\varphi}$ invariant. We may use this gauge freedom to set $\chi=h_{tt}=0$, that is to say, we set $u_0=H_0=0$ in \Eq{eq:polar}. For propagating ($\omega \neq 0$) modes, this choice completely fixes the gauge.

We find that there are only a priori five components of the equations of motion that do not vanish automatically: $\delta \hat E_t^{\;\;t}$, $\delta \hat E_z^{\;\;z}$, $\delta \hat E_t^{\;\;z}$, $\delta\hat E_\theta^{\;\;\theta} = \delta\hat E_\varphi^{\;\;\varphi}$, and $\delta \hat S$.
First, solving $\delta \hat S=0$, which is algebraic for $v(z)$, we set 
\be 
\begin{aligned}
v(z) &= - \frac{m(1-\lambda^2)^2 }{\sqrt{2}\sqrt{1+\lambda^2}} \left\{ \frac{\lambda^2}{1-\lambda^2} z H_2'(z) + \left[ -\frac{\lambda^2(1+\lambda^2)}{(1-\lambda^2)^2} + \frac{1}{2}\omega^2 z^2\right] H_2(z) \right. \\
&\hspace{33mm} + z^2 K''(z) +\frac{1-3\lambda^2}{1-\lambda^2} z K'(z) +  \left(-\frac{1+\lambda^2}{1-\lambda^2} + \omega^2 z^2\right) K(z) \\
& \hspace{33mm} \left. + \frac{i \omega z^2(1-\lambda^2)}{(1+\lambda^2)^4 m^2}\left[(1-\lambda^2) z^2 H_1'(z) + 2(1-2\lambda^2) z H_1(z)\right]\right\}.
\end{aligned}
\ee
Next, we solve for $H_{2}(z)$ via $\delta\hat E_t^{\;\;z}=0$, yielding
\be 
H_{2}(z)=-\frac{1-\lambda^2}{\lambda^{2}}\left[K(z)+ zK'(z)\right],
\ee
which we also set henceforth. 
Let us rescale to our barred variables $\bar{K}(z)=z^{-\frac{\lambda^{2}}{1-\lambda^2}}K(z)$ and $\bar{H}_{1}(z)=z^{-\frac{\lambda^{2}}{1-\lambda^2}}H_{1}(z)$. 
From $\delta \hat E_t^{\;\;t} -\delta\hat E_z^{\;\;z}=0$, we fix $\bar{H}_{1}(z)$:
\be 
\begin{aligned}
	\bar{H}_{1}(z) & =-\frac{im^{2}(1+\lambda^{2})^{4}\omega}{2\lambda^{2}(1-\lambda^2) z}\,\bar{K}(z).
\end{aligned}
\ee
We now have that $\delta\hat E_t^{\;\;t} - \delta E_z^{\;\;z}$, $\delta\hat E_t^{\;\;z}$, $\delta\hat E_\theta^{\;\;\theta}$, and $\delta\hat S$ all identically vanish, so the only remaining equation of motion is $\delta \hat E_t^{\;\;t} = 0$, which turns out to be entirely in terms of $\bar{K}$:
\be 
\bar{K}''(z)+\omega^{2}\bar{K}(z)-\frac{(2+\lambda^{2})(1+2\lambda^{2})}{(1-\lambda^2)^{2}z^{2}}\bar{K}(z)=0.
\ee
We note that the coefficient of the $1/z^{2}$ potential is precisely $g_+^{\rm pol}$ in \Eq{eq:gs} evaluated at $\ell=0$. 
It makes sense that, out of the $g^{\rm pol}_{\pm}$ modes, only $K$ would survive, since $u_{4}$ disappeared from \Eq{eq:polar} at $\ell=0$ and was replaced by $v$, which is algebraically fixed. Meanwhile, it is also understandable that the $g_0^{\rm pol}$ mode will not appear for $\ell=0$, since obtaining that mode required rescalings of the equations of motion in \Eq{eq:betahats} that become singular at $\ell=0$. 
Thus, we have found that there exists a single propagating scalar mode, and it is polar.
Notice that while it is natural to associate the scalar $\ell=0$ wave with the dilaton, it is $K$ that is actually gauge invariant, and our gauge choice to set the dilaton perturbation $u_0$ to zero means that all of the information in the scalar wave solution has been transferred to the above equation for $K$.

\section{Discussion}\label{sec:discussion}

We now examine several special topics, namely, the functional form of the propagating solutions, the $\lambda=\sqrt{3}$ theory and its relation to KK reduction of five-dimensional gravity, and the generalization of the $\lambda=0$ case to dyonic charge.

\subsection{Propagating solutions}
We have shown that the propagating linear modes around the ${\rm AdS}_2 \times S^2$ solution in Einstein-Maxwell-dilaton theory can be arranged into five towers of states indexed by angular momentum, with $\propto 1/z^2$ potentials that encode the mass term for an effective free massive field in an ${\rm AdS}_2$ background.
That is, the Regge-Wheeler-Zerilli problem in the ${\rm AdS}_2 \times S^2$ space results in modes that all satisfy a master equation given by the one-dimensional time-independent Schr\"odinger equation of conformal quantum mechanics~\eqref{eq:conformalQM}.
The solutions to the Schr\"odinger equation with $\propto1/z^{2}$ potential have many peculiar properties in the attractive case corresponding to $g<0$~\cite{EssinGriffiths}. 
However, we have found that the modes are all nontachyonic, with $g > 0$ in \Eq{eq:gs}.
The general solution to \Eq{eq:conformalQM} for nonzero $\omega$ is given by
\be 
\psi(z)=\sqrt{\omega z}\left[c_{+} H^{+}_{h-\frac{1}{2}}(\omega z)+c_{-}H^{-}_{h-\frac{1}{2}}(\omega z)\right]
\ee
for arbitrary constants $c_\pm$, where $h$ is the conformal weight corresponding to $g$ in \Eq{eq:weight} and $H^{\pm}_n(z) = J_n(z) \pm i Y_n(z)$ are the Hankel functions.\footnote{Here, $J_n(z)$ and $Y_n(z)$ are the Bessel functions of the first and second kinds, respectively. While the Hankel functions of the first and second kinds are often written as $H^{(1)}_n(z) = H^+_n(z)$ and $H^{(2)}_n(z) = H^-_n(z)$, we find the $\pm$ superscript notation more convenient.}

Considering the boundary conditions, let us focus on the $\lambda < 1$ case, for which the horizon of the original GHS black hole is at $z\rightarrow \infty$, while the AdS boundary at $z = 0$ corresponds to the region where the near-horizon throat joins the asymptotically flat geometry.
In the $z\rightarrow\infty$ limit, the solution asymptotes to a simple plane wave traveling up or down the black hole throat:
\be 
\psi(z)\stackrel{z\rightarrow\infty}{\longrightarrow}\sqrt{\frac{2}{\pi}}\left[c_{+}e^{i\left(\omega z-\frac{\pi h}{2}\right)}+c_{-}e^{-i\left(\omega z-\frac{\pi h}{2}\right)}\right].
\ee
Since the time dependence in our convention goes like $e^{i\omega t}$, with $\omega > 0$ the coefficient $c_+$ corresponds to an outgoing solution, while $c_-$ corresponds to an ingoing solution.
Near the $z=0$ boundary, we have:
\be 
\psi(z)\stackrel{z\rightarrow0}{\longrightarrow}  \frac{\sqrt{2}}{\Gamma\left(h{+}\frac{1}{2}\right)}\left[c_+ {+} c_- - i (c_+ {-} c_-) \tan \pi h \right]\left(\frac{\omega z}{2}\right)^{h} -\frac{i\sqrt{2}(c_{+}{-}c_{-})}{\pi}\Gamma\left(h{-}\tfrac{1}{2}\right)\left(\frac{\omega z}{2}\right)^{1-h}.\label{eq:psismallz}
\ee 
Notice that the boundary conditions for all the various modes are the same because they all obey the same master equation~\eqref{eq:conformalQM}. This implies in particular that for the mode most closely associated with the dilaton, $\psi=w_0$, the behavior of the perturbation does not in any way alter the background dilaton's logarithmic divergence near the Poincar\'e horizon of ${\rm AdS}_2$.
We note that the conformal weights of all five towers of modes (for all $\ell$ to which they apply) satisfy $h\geq 2$.
Typical AdS/CFT boundary conditions would be to impose Dirichlet or Neumann conditions, killing one of the modes that behave as $z^h$ or $z^{1-h}$ (or at least imposing mixed boundary conditions that would imply zero flux at the AdS boundary). 
However, it is well known that, for ${\rm AdS}_2$, back reaction destroys any such asymptotically ${\rm AdS}$ boundary conditions \cite{Maldacena:1998uz, Galloway:2018dak}. 
On the other hand, if we embed our ${\rm AdS}_2 \times S^2$ in a larger geometry---as the near-horizon limit of extreme GHS in string frame---then the solution is modified near the ${\rm AdS_2}$ boundary, and it becomes possible and indeed necessary to consider ``leaky'' boundary conditions with nonzero net flux near the would-be ${\rm AdS_2}$ boundary.
This was discussed in the context of the Reissner-Nordstr\"om black hole in Refs.~\cite{Porfyriadis:2018yag,Porfyriadis:2018jlw}.
Such an exploration of deviations away from the near-horizon limit would be necessary, for example, if we wanted to impose the requisite boundary conditions to derive the eigenfrequencies of the quasinormal modes.
We leave further exploration of such questions to future work.

\subsection{$\lambda = \sqrt{3}$ and Kaluza-Klein}
As noted previously, the Einstein-Maxwell-dilaton action in Eq.~\eqref{eq:action} with $\lambda=\sqrt{3}$ arises from the KK compactification of five-dimensional Einstein gravity on a circle. 
Let us see how this comes about. We start with the $D=5$ Einstein-Hilbert action:
\be 
{\cal S}=\frac{1}{2\kappa_{5}^{2}}\int{\rm d}^{4}x\,{\rm d}y\sqrt{-\det g_{AB}}\,^{(5)}\! R,
\ee
where we parameterize the five-dimensional metric in terms of coordinates $x^{A}$ and reserve Greek indices for the four-dimensional spacetime:
\be 
g_{AB}=\left(\begin{array}{cc}
\mathring {g}_{\mu\nu}+2\Phi^{2}A_{\mu}A_{\nu} & \sqrt{2}\Phi^{2}A_{\nu}\\
\sqrt{2} \Phi^{2}A_{\mu} & \Phi^{2}
\end{array}\right).\label{eq:gAB5d}
\ee 
One obtains a four-dimensional action upon cylindrical compactification, assuming all fields are $y$-independent and defining $F_{\mu\nu}$ as the field strength associated with the vector $A_\mu$~\cite{Williams},
\be 
{\cal S}=\frac{L}{2\kappa_{5}^{2}}\int{\rm d}^{4}x\sqrt{-\mathring{g}}\left(\Phi\mathring{R}-\frac{1}{2}\Phi^{3}F_{\mu\nu}F^{\mu\nu}\right),
\ee
writing $L$ for the circumference of the cylinder in the $y$-coordinate and $\mathring R$ for the Ricci curvature of $\mathring g_{\mu\nu}$, and where all indices are raised using $\mathring{g}^{\mu\nu}$. Defining a new metric $g_{\mu\nu}=\Phi\mathring{g}_{\mu\nu}$ and $\Phi=e^{-2\phi/\sqrt{3}}$, and setting units where $L/2\kappa_{5}^{2}=1$, the Lagrangian then becomes
\be 
{\cal L}=R-2(\nabla\phi)^{2}-\frac{1}{2}e^{-2\sqrt{3}\phi}F_{\mu\nu}F^{\mu\nu},
\ee
i.e., the action in Eq.~\eqref{eq:action}, with $\lambda=\sqrt{3}$, and we see that $g_{\mu\nu}$ is simply the metric for the corresponding solution in Einstein frame.

Thus, for $\lambda = \sqrt{3}$, our ${\rm AdS}_{2}\times S^{2}$ solution should be a compactification of some vacuum solution in five-dimensional Einstein gravity.
Let us explore this vacuum construction.
First, going from the string-frame metric of Eq.~\eqref{eq:metstring}, in the case $\lambda=\sqrt{3}$,
\be 
({\rm d}s^{2})_{{\rm string}}=\frac{4r_{0}^{2}}{z^{2}}(-{\rm d}t^{2}+{\rm d}z^{2})+r_{0}^{2}{\rm d}\Omega^{2},
\ee
to four-dimensional Einstein frame, $(g_{\mu\nu})_{{\rm Einstein}}=e^{-2\lambda\phi}(g_{\mu\nu})_{{\rm string}}$ with $\phi=\frac{\lambda}{1-\lambda^2}\log z$, we have
\be 
({\rm d}s^{2})_{{\rm Einstein}}=r_0^2 [4z(-{\rm d}t^{2}+{\rm d}z^{2})+z^3 {\rm d}\Omega^{2}].
\ee
Meanwhile, a choice of gauge field corresponding to $F$ as given in Eq.~\eqref{eq:metstring} is 
\be 
A = \frac{r_0}{\sqrt{2}} (1-\cos \theta) \,{\rm d}\varphi,
\ee
defined everywhere except for the Dirac string at $\theta\,{=}\,\pi$.
Translating this into the five-dimensional metric using Eq.~\eqref{eq:gAB5d} and defining dimensionful $(T,Z) = 2r_0 (t,z)$, after some algebraic rearrangement we have:
\be 
{\rm d}s^{2}=-{\rm d}T^{2}+{\rm d}Z^{2} +\frac{Z^2}{4} \left\{\left[\frac{{\rm d}y}{r_0} + (1-\cos\theta){\rm d}\varphi \right]^2 + {\rm d}\theta^2 + \sin^2\theta\, {\rm d}\varphi^2 \right\}.\label{eq:5dtest}
\ee
Explicitly evaluating the Riemann tensor for the metric in Eq.~\eqref{eq:5dtest}, one finds that this is indeed a solution to the five-dimensional vacuum Einstein equations. What sort of five-dimensional geometry is this? We recognize the term in braces $\propto Z^2$ in \Eq{eq:5dtest} as the metric for the Hopf fibration of $S^3$, with the compact dimension $y$ playing the role of the Hopf fiber.
To view this another way, let us define new coordinates $\vartheta = \theta/2$, $\zeta_1 = \varphi + (y/2r_0)$, and $\zeta_2 = y/2r_0$, so that Eq.~\eqref{eq:5dtest} becomes
\be 
{\rm d}s^{2}=-{\rm d}T^{2}+{\rm d}Z^{2}+Z^{2}\left({\rm d}\vartheta^{2}+\sin^{2}\vartheta\,{\rm d}\zeta_1^{2}+\cos^{2}\vartheta\,{\rm d}\zeta_{2}^{2}\right).\label{eq:Hopf}
\ee
In this form, $(\vartheta,\zeta_1,\zeta_2)$ define Hopf coordinates, where $\vartheta$ runs from $0$ to $\pi/2$ and $\zeta_{1,2}$ each run from $0$ to $2\pi$.
At each fixed $\vartheta$, $\zeta_{1,2}$ parameterize a torus.
Avoiding a conical singularity requires that $y/r_0$ be periodic in  $4\pi$, that is, we must identify $L$ as $4\pi r_0$.
The flat geometry in Eqs.~\eqref{eq:5dtest} and \eqref{eq:Hopf} is the metric near the KK monopole~\cite{Horowitz:2011cq,Gross:1983hb,Sorkin:1983ns}, which in the four-dimensional effective theory is described in string frame at short distance by the $\lambda = \sqrt{3}$ case of our ${\rm AdS}_2 \times S^2$ solution.

\subsection{Dyonic Reissner-Nordstr\"om}\label{sec:RN}

The magnetic Bertotti-Robinson geometry obtained via the near-horizon limit of the extremal magnetic Reissner-Nordstr\"om black hole has a spectrum that we can read off from the $\lambda=0$ case of the results in previous sections.
The dilatonic polar mode $w_0$ must be discarded, leaving us with two polar and two axial towers of massive modes for general $\ell \geq 2$.
We find that there is a striking ${\mathbb Z}_2$ symmetry in the spectra associated with exchanging the axial and polar modes:
\be
\begin{aligned}
g_+^{\rm pol}|_{\lambda = 0} = g_+^{\rm ax}|_{\lambda = 0} &= (\ell+2)(\ell+1) \\
g_-^{\rm pol}|_{\lambda = 0} = g_-^{\rm ax}|_{\lambda = 0} &= \ell(\ell-1).
\end{aligned} \label{eq:RNmag}
\ee
Unlike the dilatonic general-$\lambda$ case, the extremal electric Reissner-Nordstr\"om black hole---or more generally, an extremal dyonic Reissner-Nordstr\"om black hole with arbitrary electric and magnetic charges---exhibits a near-horizon Bertotti-Robinson geometry:
\be
\begin{aligned}
{\rm d}s^2 &= \frac{r_0^2}{z^2} (-{\rm d}t^2 + {\rm d}z^2) + r_0^2 {\rm d}\Omega^2 \\
F &= \frac{q}{z^2}{\rm d}t\wedge {\rm d}z + p \sin\theta\,{\rm d}\theta\wedge {\rm d}\varphi,
\end{aligned}
\ee
where $q^2 + p^2 = 2r_0^2$. 
We define the ratio between the electric and magnetic charges as a dyonic angle, $p/q = \tan \eta$.
This naturally suggests the question: What is the spectrum of propagating modes in this dyonic ${\rm AdS}_2 \times S^2$ solution in Einstein-Maxwell theory?
While the pure magnetic case gives us the answer for $\eta = \pi/2$, let us now investigate this question for all other values of $\eta$.

Since the background will in general be of indefinite parity in the gauge field due to the dyonic charge, let us take the ansatz for the photon to be a combination of those in Eqs.~\eqref{eq:polar} and \eqref{eq:axial}:
\be 
a_\mu = \left(u_1(z)P_\ell(\cos\theta),\;\;u_2(z)P_\ell(\cos\theta),\;\;0,\;\; u_4(z)\sin\theta \frac{{\rm d}}{{\rm d}\theta}P_\ell(\cos\theta)\right)\times e^{i\omega t}.\label{eq:photongeneral}
\ee
As noted in \Sec{sec:linearized}, we are able to set $u_3(z) = 0$ by a gauge choice.
Despite our more general form for the gauge field perturbation, it will be useful to organize the graviton and dilaton modes by parity just as in Eqs.~\eqref{eq:polar} and \eqref{eq:axial}.
Let us start with the case of the axial graviton perturbation~\eqref{eq:axial}. We find that $\delta\hat E_t^{\;\;t} = 0$ fixes $u_2(z)$,
\be
i\omega u_2(z) = u_1'(z) + \frac{\ell(\ell+1)\tan\eta}{z^2} u_4(z),
\ee
while $\delta\hat E_\theta^{\;\;\varphi} = 0$ fixes $h_0(z)$ via $i\omega h_0(z) =  h_1'(z)$.
Defining components of the equations of motion via the $\alpha_i$ and $\beta_i$ in Eqs.~\eqref{eq:alphas} and \eqref{eq:betas}, with $\lambda$ set to zero, we find that the remaining components that do not already identically vanish are $\alpha_{0,1,2,3,4}$ and $\beta_{0,3,4}$.
Among these eight equations, five relations are identically satisfied:
\be
\begin{aligned}
2\sqrt{2} \alpha_0 \cos\eta + \frac{\beta_4}{z} &=0 \\
2\sqrt{2}\alpha_1 \cos\eta + i \omega \beta_3 &=0 \\
i\omega \alpha_0 + \frac{{\rm d}\alpha_1}{{\rm d}z} - \frac{\ell(\ell+1)}{z}\alpha_2 &= 0 \\
i\omega \alpha_1 + \frac{{\rm d}\alpha_0}{{\rm d} z} - \ell(\ell+1) \beta_0 \tan\eta &= 0 \\
2\sqrt{2} r_0 \alpha_2 \sin\eta - 2i\alpha_3 - z \frac{{\rm d}\alpha_4}{{\rm d}z} &= 0.
\end{aligned} 
\ee
It will therefore suffice to solve three independent equations. From $\alpha_0 = 0$, we have an expression for $u_1(z)$ in terms of $h_1$ and $u_4$:
\be 
u_1(z)  = \frac{\sqrt{2}}{i\omega r_0}h_1'(z)\sin\eta - u_4'(z)\tan\eta .
\ee
The two remaining undetermined functions satisfy a set of coupled second-order equations given by $2\sqrt{2}r_0 \alpha_1 \sin\eta  -\ell(\ell+1)\alpha_4 = 0$ and $2\sqrt{2}r_0 \alpha_1 \sin\eta  - \ell(\ell+1)\alpha_4 + \frac{i\omega \ell(\ell+1)r_0 z^2}{\sqrt{2}\cos\eta} \beta_0= 0$, which we can write as a matrix equation~\eqref{eq:matrixeq}, but where our constant matrix is now
\be 
\mathbf{A}=\left(\begin{array}{cc}
(\ell+2)(\ell-1) & 2\sqrt{2}i \sec \eta \\
-\sqrt{2}i(\ell+2)(\ell-1)\cos\eta & \ell^2 + \ell + 4
\end{array}\right)
\ee
and our vector is
\be 
\vec{\mathbf{x}}=\left(\begin{array}{c}
{h}_{1}(z)\\
r_0 \omega {u}_{4}(z)
\end{array}\right).
\ee 
Diagonalizing, we find the propagating modes in the axial case,
\be 
\begin{aligned}
y_+(z) &= -i(\ell-1)h_1(z)\cos\eta + \sqrt{2}r_0 \omega u_4(z) \\
y_-(z) &= i(\ell+2) h_1(z)\cos\eta + \sqrt{2}r_0 \omega u_4(z),
\end{aligned}
\ee
which satisfy the master equations $y_\pm''(z) + \omega^2 y_\pm (z) = g^{\rm ax,RN}_\pm y_\pm(z)/z^2$, where 
\be
\begin{aligned}
g_-^{\rm ax,RN} &= \ell(\ell-1) \\
g_+^{\rm ax,RN} &= (\ell+2)(\ell+1),
\end{aligned} \label{eq:RNax}
\ee
corresponding to weights $h_-^{\rm ax,RN} = \ell$ and $h_+^{\rm ax,RN} = \ell+2$.

Let us now turn to the case of the polar graviton and dilaton perturbations in \Eq{eq:polar}, but with the general photon perturbation of \Eq{eq:photongeneral}.
We find that setting $\delta\hat E_\theta^{\;\;\theta} - \delta\hat E_\varphi^{\;\;\varphi} = 0$ implies $H_2(z) = H_0(z)$, while fixing $\delta\hat E_t^{\;\;z} = 0$ yields
\be
H_1(z) = \frac{2i\omega r_0^2}{\ell(\ell+1)z}[K(z) + z K'(z)]. 
\ee
Enforcing $\delta\hat E_z^{\;\;\varphi} = 0$ gives us $u_4$ in terms of $u_2$ via $i\omega u_4(z) =  u_2(z) \tan\eta$, while $\delta\hat E_t^{\;\;\varphi} = 0$ fixes $u_1$ in terms of $u_2$, $i\omega u_1(z)  = u_2'(z)$.
For the remaining equations of motion, defining the $\alpha_i$ and $\beta_i$ as in Eqs.~\eqref{eq:alphas} and \eqref{eq:betas} with $\lambda = 0$, explicit computation shows that $\alpha_{0,1}$ and $\beta_{0,1,2,3,4,5}$ do not identically vanish.
We can use $\beta_1 = 0$ to set $H_0$,
\be
H_0(z) = \frac{z^2}{\ell(\ell+1)}\left[K''(z) + \frac{2}{z}K'(z) - \omega^2 K(z) \right],
\ee
leaving us with two unfixed functions, $K$ and $u_4$. Of the seven remaining equations, we find that five relations are identically satisfied,
\be
\begin{aligned}
i\omega\alpha_0 + \frac{{\rm d}\alpha_1}{{\rm d}z} &= 0\\
i\omega z^2 \beta_0  - \alpha_1 \tan\eta &= 0 \\
z\frac{{\rm d}\beta_3}{{\rm d}z} + \beta_4 &= 0 \\
\frac{2\sqrt{2}i\sin\eta\tan\eta}{\omega z^2} \alpha_1 + r_0 \beta_2 +\frac{1}{\ell(\ell+1)}\left(\frac{{\rm d}^2}{{\rm d}z^2} + \omega^2 \right)\beta_3 &= 0 \\
\frac{2\sqrt{2}\cos\eta}{i\omega}\alpha_1 + \beta_3 - \frac{\beta_5}{\ell(\ell+1)} &= 0,
\end{aligned} 
\ee
so it suffices to require $\alpha_1 = \beta_3 = 0$.
These expressions give a pair of coupled second-order equations for $K$ and $u_4$ of the form in \Eq{eq:matrixeq}, with a new constant matrix,
\be
\mathbf{A}=\left(\begin{array}{cc}
\ell^2 + \ell + 2 & -2\sqrt{2}i \ell(\ell+1)\sec\eta\\
\sqrt{2}i\cos\eta & \ell(\ell+1)
\end{array}\right),
\ee
and new vector containing the photon/graviton degrees of freedom,
\be 
\vec{\mathbf{x}}=\left(\begin{array}{c}
K(z)\\
u_2(z)/\omega r_0
\end{array}\right).
\ee
Diagonalizing, we find the propagating modes in the polar case,
\be
\begin{aligned}
w_-(z) &= K(z) \cos \eta +\frac{\sqrt{2}i(\ell+1)}{r_0 \omega} u_2(z) \\
w_+(z) &= K(z) \cos \eta - \frac{\sqrt{2}i\ell}{r_0 \omega} u_2(z),
\end{aligned} 
\ee
which satisfy the master equations $w_\pm''(z) + \omega^2 w_\pm(z) = g_\pm^{\rm pol,RN}w_\pm(z)/z^2$, where 
\be
\begin{aligned}
g_-^{\rm pol,RN} &= \ell(\ell-1) \\
g_+^{\rm pol,RN} &= (\ell+2)(\ell+1),
\end{aligned} \label{eq:RNpol}
\ee
corresponding to weights $h_-^{\rm pol,RN} = \ell$ and $h_+^{\rm pol,RN} = \ell+2$.

We find that the near-horizon spectrum of the dyonic black hole exhibits a ${\rm U}(1) \times \mathbb{Z}_2$ symmetry, evident at the level of the effective potential.
The ${\rm U}(1)$ symmetry is reflected in the fact that all values of $g$ are independent of the dyon angle, while the spectrum further possesses a $\mathbb{Z}_2$ symmetry associated with exchanging the axial and polar modes, $g_+^{\rm pol,RN} = g_+^{\rm ax,RN} = (\ell+2)(\ell+1)$ and $g_-^{\rm pol,RN} = g_-^{\rm ax,RN} = \ell(\ell-1)$, independent of $\eta$, as shown in Eqs.~\eqref{eq:RNmag}, \eqref{eq:RNax}, and \eqref{eq:RNpol}.
This $\mathbb{Z}_2$ symmetry is a consequence of the emergent supersymmetry of the perturbations around a Reissner-Nordstr\"om black hole discovered by Chandrasekhar~\cite{Chandrasekhar}; the potentials for polar and axial perturbations in the Reissner-Nordstr\"om background are superpartners descending from a single superpotential, and as a consequence the quasinormal modes have the same spectrum of eigenfrequencies $\omega$.\footnote{The values of $\omega$ for the quasinormal modes have boundary conditions imposed by features of the geometry away from the near-horizon ${\rm AdS}_2$ limit, and are often found numerically. This spectrum is not to be confused with the spectrum we have found, giving the effective {\it masses} of the modes propagating in the ${\rm AdS}_2$ background, with arbitrary values of $\omega$.}
What we have shown in this subsection is somewhat stronger: the master equations describing the conformal potentials governing the modes are in fact {\it identical} in the  near-horizon limit, as a consequence of the conformal symmetry of the ${\rm AdS}_2$ background.

\section{Conclusions}\label{sec:conclusions}

In this work, we have identified a new ${\rm AdS}_2 \times S^2$ construction in Einstein-Maxwell-dilaton gravity, which can be obtained via a near-horizon limit of the GHS black hole in string frame.
This solution is a compelling object of study, due in particular to the customizable ratio it exhibits between the AdS and compact length scales controlled by the dilaton coupling.

After presenting the solution in \Sec{sec:AdS}, we turned to an investigation of the perturbation theory of  the graviton, photon, and dilaton degrees of freedom in this background.
We derived the effective potential for the independently propagating modes, finding that they all obey the conformal Schr\"odinger equation with couplings given in \Sec{sec:summary}.
The details of the calculation are presented in Secs.~\ref{sec:linearized}, \ref{sec:axial}, and \ref{sec:polar}, with the special small-$\ell$ cases handled in Secs.~\ref{sec:l1} and \ref{sec:l0}.
In \Sec{sec:discussion}, we discussed several special topics, including the explicit form of the propagating solutions as Hankel functions, the case of $\lambda = \sqrt{3}$ and its connection to five-dimensional general relativity under compactification via Hopf fibration, and a generalization to derive the spectrum in dyonically charged Bertotti-Robinson space.

This paper leaves several interesting directions for future work.
An understanding of the full set of nonpropagating perturbations, and any relation they bear to the ${\rm SL}(2)$ symmetry of the metric, is deserving of study.
In particular, such solutions would include the analogues of the anabasis, elucidated for Reissner-Nordstr\"om in Ref.~\cite{Hadar:2020kry}, that enhance the near-horizon ${\rm AdS}_2\times S^2$ geometry by the addition of leading-order corrections away from the black hole throat in the extremal GHS spacetime.
Such an investigation would be necessary in order to find the boundary conditions that fix the frequencies of quasinormal modes.
For the propagating solutions, the connection with waves in the far asymptotically flat region of GHS is of great interest as well. The solution to this connection problem, along the lines of Refs.~\cite{Porfyriadis:2018yag, Porfyriadis:2018jlw} for Reissner-Nordstr\"om, is left for future work; a good starting point could be the decoupled equations for GHS perturbations first identified in Ref.~\cite{Holzhey:1991bx}.
The partial breaking of ${\rm SL}(2)$ symmetry in our ${\rm AdS}_2 \times S^2$ solution also suggests possible implications of this feature for the attractor mechanism and entropy function formalism, which generally assume the presence of full  ${\rm SL}(2)$ symmetry in the near-horizon limit of extreme black holes~\cite{Sen:2007qy}.

More broadly, the perturbative analysis conducted in this paper will be necessary for any holographic construction based on our ${\rm AdS}_2 \times S^2$ solution.
As we have discussed, the couplings in the $1/z^2$ potential can be thought of as the effective masses of the states propagating in the ${\rm AdS}_2$ background and give us the conformal weights of boundary operators dual to these modes.
This characterization of the set of conformal weights is a crucial ingredient in constructing a boundary theory.
We leave the quest to build a theory matching this spectrum to future investigations.

\vspace{12mm}
 
\begin{center} 
{\bf Acknowledgments}
\end{center}
\noindent 
We thank Cliff Cheung, Xi Dong, Gary Horowitz, Yu-tin Huang, Don Marolf, Alexey Milekhin, and Andy Strominger for useful discussions and comments.
A.P.P. is supported by the Black Hole Initiative at Harvard University, which is funded by grants from the John Templeton Foundation and the Gordon and Betty Moore Foundation.
G.N.R. is supported at the Kavli Institute for Theoretical Physics by the Simons Foundation (Grant~No.~216179) and the National Science Foundation (Grant~No.~NSF PHY-1748958) and at the University of California, Santa Barbara by the Fundamental Physics Fellowship.

\pagebreak

\bibliographystyle{utphys-modified}
\bibliography{AdS2GHS}

\end{document}